\documentclass[numberedappendix]{emulateapj}
\pdfoutput=1
\usepackage{apjfonts}
\usepackage{amssymb}
\usepackage{amsmath}
\usepackage{subfigure}
\usepackage{multirow}

\shorttitle{}
\shortauthors{Mao et al.}

\begin{document}
\title{New Constraints on the Galactic Halo Magnetic Field using Rotation Measures of Extragalactic Sources Towards the Outer Galaxy}
\author{S. A. Mao,\altaffilmark{1,2,3,4} 
N. M. McClure-Griffiths,\altaffilmark{2}
B. M. Gaensler,\altaffilmark{5}
J. C. Brown, \altaffilmark{6}
C. L. van Eck, \altaffilmark{6}
M. Haverkorn,\altaffilmark{7}
P. P. Kronberg, \altaffilmark{8,9}
J. M. Stil,\altaffilmark{6}
A. Shukurov,\altaffilmark{10}
A. R. Taylor \altaffilmark{6}
}
\altaffiltext{1}{Harvard-Smithsonian Center for Astrophysics, Cambridge, MA 02138; mao@astro.wisc.edu}
\altaffiltext{2}{Australia Telescope National Facility, CSIRO, Epping, NSW 1710, Australia}
\altaffiltext{3}{Jansky Fellow, National Radio Astronomy Observatory, P.O. Box O, Socorro, NM 87801}
\altaffiltext{4}{Department of Astronomy, University of Wisconsin, Madison, WI 53706}
\altaffiltext{5}{Sydney Institute for Astronomy, School of Physics, The University of Sydney, NSW 2006, Australia}
\altaffiltext{6}{Department of Physics and Astronomy, and Institute for Space Imaging Science, University of Calgary, AB, Canada}
\altaffiltext{7}{Department of Astrophysics, Radboud University, P.O. Box 9010, 6500 GL Nijmegen, The Netherlands}
\altaffiltext{8}{Los Alamos National Laboratory, P.O. Box 1663, Los Alamos, NM 87545}
\altaffiltext{9}{Department of Physics, University of Toronto, 60 St. George Street, Toronto, M5S 1A7, Canada}
\altaffiltext{10}{School of Mathematics and Statistics, University of Newcastle, Newcastle upon Tyne, NE1 7RU, UK}

\begin{abstract}
We present a study of the Milky Way disk and halo magnetic field, determined from observations of Faraday rotation measure (RM) towards 641 polarized extragalactic radio sources in the Galactic longitude range 100$^\circ$-117$^\circ$, within 30$^\circ$ of the Galactic plane. For $|$$b$$|$ $<$ 15$^\circ$, we observe a symmetric RM distribution about the Galactic plane. This is consistent with a disk field in the Perseus arm of even parity across the Galactic mid-plane. In the range 15$^\circ$$<$$|$$b$$|$ $<$ 30$^\circ$, we find median rotation measures of $-$15$\pm$4 rad m$^{-2}$ and $-$62$\pm$5 rad m$^{-2}$ in the northern and southern Galactic hemispheres, respectively. If the RM distribution is a signature of the large-scale field parallel to the Galactic plane, this suggests that the halo magnetic field toward the outer Galaxy does not reverse direction across the mid-plane. The variation of RM as a function of Galactic latitude in this longitude range is such that RMs become more negative at larger $|b|$. This is consistent with an azimuthal magnetic field of strength 2 $\mu$G (7$\mu$G) at a height 0.8-2 kpc above (below) the Galactic plane between the local and the Perseus spiral arm. We propose that the Milky Way could possess spiral-like halo magnetic fields similar to those observed in M51.

\end{abstract}


\section{Introduction}
\label{vlasection:introduction}

Magnetism is an important component of the interstellar medium (ISM): it contributes to gas pressure to balance it against gravity, achieving hydrostatic equilibrium; it helps to remove angular momentum from a collapsing molecular cloud, allowing star formation to proceed;  and it also confines and deflects cosmic rays. Yet, we still do not understand the details of processes that generate large-scale coherent galactic magnetic fields, especially those in regions outside the galactic disk.

The existence of large-scale magnetic fields in galactic disks can be explained by the classical mean-field dynamo -- the amplification of magnetic fields by small-scale turbulent motion (the $\alpha$-effect) and differential rotation (the $\omega$-effect) on a time scale of 10$^9$ years \citep{shukurov2004}. According to dynamo theory, an axisymmetric magnetic field with quadrupole symmetry (even parity) with respect to the galactic mid-plane is most readily excited in the disk. This is found to be the case for the Milky Way -- its disk field has even parity across the Galactic plane \citep[e.g.,][]{frick2001,taylor2009,kronberg2011}. 

Dynamo modes excited in the thick disks or halos of galaxies can have dipolar symmetry (odd parity) with respect to the galactic disk. These modes could also be oscillatory, resulting in large-scale transient magnetic field reversals \citep{sokoloff1990,brandenburg1992,moss2010}.  Unfortunately, it is difficult  to establish the halo field symmetry with respect to the galactic plane even in the most well-studied edge-on galaxy, NGC 253. \cite{braun2010} found that a large-scale magnetic field model of an axisymmetric spiral form plus an out-of-plane quadrupole extension that could reproduce polarization patterns observed in a number of face-on galaxies. Unfortunately, the backside halos in their target galaxies were severely depolarized and the field symmetry with respect to the mid-plane could not be determined. Disk and halo fields could be of different axial geometries: for example, M 51 (inclination of $-$20$^\circ$) was found to have a bisymmetric halo field and an axisymmetric disk field \citep{fletcher2011}.

Characterizing the halo magnetic field symmetry in our own Milky Way is even more challenging because of our location within it.
Early works based on a sparsely sampled all-sky Faraday rotation measure (RM) distribution of pulsars and extragalactic sources (EGSs) have suggested a dipolar-type Galactic halo magnetic field \citep{andreassian1988,han1997}. However, using the RM of $\sim$ 1,000 extragalactic sources, \cite{mao2010} demonstrated that the Milky Way's vertical magnetic field at the location of the Sun is of neither quadrupolar nor dipolar symmetry. More recently, \cite{wolleben2010} used RMs of diffuse polarized Galactic synchrotron emission to show that the EGS RM pattern thought to be the signature of a dipolar halo field is at least partly due to a local magnetized shell. Clearly, further effort is needed to properly describe the geometry of the Galactic halo magnetic field. 

In this paper, we characterize the symmetry of the Galactic halo magnetic field by studying RMs of EGSs in the region 100$^\circ$$<$$l$$<$117$^\circ$, $|$$b$$|$$<$30$^\circ$ (towards the Perseus Arm). This study is based on existing EGS RMs in the Canadian Galactic Plane Survey (CGPS) latitude extension region (100$^\circ$$<$$l$$<$117$^\circ$, $-$3$^\circ$$<$$b$$<$+18$^\circ$, Brown et al. in prep) and a new Very Large Array (VLA) RM survey that extends the latitude coverage of the RM data to $|$$b$$|$$<$30$^\circ$. The latitude coverage of this new VLA data set enables one to study the parity of the halo field across the Galactic plane. The longitude range is ideal because it samples the magneto-ionic medium half way between the two strips of the Galactic plane that will be covered by the GALFA Continuum Transit Survey (GALFACTS) \citep{taylor2010}.

In Section~\ref{vlasection:observations}, we describe the VLA observations and data reduction. We then compute EGS RMs and present their spatial distributions in Section~\ref{vlasection:results}. In Section~\ref{vlasection:disk_symmetry}, we infer properties of the symmetric disk field towards the Perseus spiral arm using the observed RM-vs-$b$ trend within $\sim$15$^\circ$ of the Galactic plane.  In Section~\ref{vlasubsection:location} and Section~\ref{vlasubsection:enhancement}, we show that  RMs in the surveyed region are likely produced by a large-scale magnetic field in the warm ionized medium (WIM). In Section~\ref{vlasubsection:halo_models}, we illustrate that the existing halo field models cannot reproduce satisfactorily the RM-vs-$b$ trend in our survey area. We construct a simple Galactic halo field model in Section~\ref{vlasubsection:B_toy_model} that can successfully reproduce the general observed RM trend in the surveyed area.

\section{Observations and Data Reduction}
\label{vlasection:observations}

To directly measure the parity of the Milky Way's disk and halo magnetic field in a latitude strip, we extend the latitude coverage of the CGPS latitude extension region by conducting an EGS RM survey with the VLA. We construct a magnetic field model for the Milky Way to estimate the number of sources needed for such a detection. We assume the Milky Way has a symmetric 1 $\mu$G disk field with  a scale height of 1 kpc  and an antisymmetric 1 $\mu$G halo field with a scale height of 4 kpc. In addition, we use the thermal electron thick disk model derived by \cite{cl2002} as the Faraday rotating layer, for the purpose of observation planning. We find at latitudes higher than 30$^\circ$, the difference in RMs above and below the plane due to the halo field might be too small to determine its parity, hence we use $|$$b$$|$=30$^\circ$ as the latitude boundary of our observations. This halo field toy model predicts an RM difference of $\sim$ 40 rad m$^{-2}$ for $|b|$$>$10$^\circ$. Since the standard deviation of EGS RMs at $|$$b$$|$$\sim$30$^\circ$ is $\sim$ 18 rad m$^{-2}$ \citep[e.g.,][]{schnitzeler2010}, one needs approximately 15 sources in each 1$^\circ$ latitude bin to distinguish the RM trend for  $b$$>$0$^\circ$  from $b$$<$0$^\circ$, resulting in a source requirement of $\sim$ 900 EGSs within 30$^\circ$ of the Galactic plane. The CGPS latitude extension survey provides $\sim$ 300 extragalactic sources in the latitude range $-$3.5$^\circ$$<$$b$$<$+18$^\circ$ (Brown et al. in prep). Therefore, we have selected $\sim$ 600 compact extragalactic sources from  the NRAO VLA Sky Survey (NVSS) catalogue \citep{condon1998} with polarized intensity brighter than 4 mJy in the region 100$^\circ$$<$$l$$<$117$^\circ$, -30$^\circ$$<$$b$$<$-3$^\circ$ and +18$^\circ$$<$$b$$<$+30$^\circ$ as targets.

Observations of the selected polarized EGSs were conducted at the VLA on 2008 June 8th and 23rd, July 7th and 8th, August 26th and 31st, and September 1st and 2nd  in D and DnC configurations. Data were taken in two 25 MHz bands in the spectral line mode. Each band consists of seven 3.125 MHz-wide channels (with an edge channel discarded), centered at 1365 MHz and 1486 MHz, respectively. Data taken on June 8th had the higher frequency band centered at 1512 MHz. Unfortunately, these latter data were rendered unusable due to severe radio frequency interference (RFI). The two frequency bands are separated by 100 MHz to ensure precise RM determination, as the uncertainty in RM is inversely proportional to the span in the coverage in wavelength squared. Each source was observed for $\sim$ 1.6 minutes at each frequency band. Including slew times and overheads, the total observing time was $\sim$ 40 hours. Standard VLA primary calibrators 3C147 and 3C286 were observed at the beginning and the end of each observing run for absolute flux and polarization leakage calibration. Polarized sources 3C286 and 3C48 were observed for the purpose of absolute polarization angle calibration. In addition, a secondary calibrator (one of the following: 1800+784, 2005+778, 2133+826, 2355+498, 0029+349) located closest to the target EGSs was observed every hour to obtain time-dependent antenna gains.

We converted the raw data into UVFITS format using the task FITTP in AIPS in order to carry out data calibration, and then make images using the MIRIAD package \citep[][]{sault1995}\footnotemark[1]\footnotetext[1]{See also http://www.ph.unimelb.edu.au/$\sim$rsault/software/miriad/vla-polarimetry.html for specifics of reducing VLA data in MIRIAD.}. Flux densities of the sources were calibrated to 3C147, whose Stokes I value at 1365 MHz and 1486 MHz is  22.5 Jy and 21.2 Jy, respectively. The absolute polarization angles of the sources were calibrated to either 3C286, whose angle at 1365 MHz and 1486 MHz is  +33$^\circ$, measured from North through East, with a rotation measure\footnotemark[2]\footnotetext[2]{http://www.vla.nrao.edu/astro/calib/manual/}  of 0 rad m$^{-2}$ or 3C48, whose angle is tied to that of 3C286: $-$55$^\circ$ at 1365 MHz and $-$30$^\circ$ at 1486 MHz. Time-dependent antenna gains, polarization leakages and absolute phases of the 14 spectral channels were calibrated individually.  For each EGS, images of Stokes $Q$ and $U$ were made in each of the 14 3.125-MHz wide frequency channel using natural weighting to maximize sensitivity. This results in channels maps each with a sensitivity of roughly 0.4 mJy/beam. We excluded the shortest baselines ($uv$ distance $<$ 0.5 k$\lambda$) to avoid imaging extended Galactic synchrotron emission. The Stokes $Q$ and $U$ channel maps were then deconvolved and restored to the resolution at 1356 MHz ($\sim$1'). A linearly polarized intensity (PI) map corrected for positive bias was made for each source. The brightest polarized pixel of each EGS was identified and its Stokes $Q$ and $U$ values across the frequency band were extracted for RM determination. 

This work also makes use of 339 EGS RMs in the CGPS latitude extension region (100$^\circ$$<$$l$$<$117$^\circ$, $-$3.5$^\circ$$<$$b$$<$+18$^\circ$, Brown et al. in prep). These EGSs have a mean total intensity of 81 mJy/beam and a mean polarized fraction of 5\%. RMs are determined using Stokes $Q$ and $U$ measured in four 7.5-MHz channels centered at 1420 MHz. The typical RM error is 15 rad m$^{-2}$. 

\section{Results}
\label{vlasection:results}
\subsection{RM Computation}
\label{vlasubsection:rm_computation}

Faraday rotation is a birefringence effect \citep[see e.g. Section 4.3 of][for more details]{mao2010}. We have computed rotation measures using RM Synthesis and RMCLEAN \citep{brentjens2005,heald2009}, following the algorithm presented by \cite{mao2010}. The RM uncertainty is computed by directly measuring the noise in the real and imaginary parts of the RM spectrum. Our VLA observing frequency setup results in RM spectra with FWHM $\sim$ 400 rad m$^{-2}$ and a maximum detectable RM of 10$^{4}$ rad m$^{-2}$. 

Recently, \cite{farnsworth2011} demonstrated that using RM synthesis alone might not be sufficient to determine the underlying Faraday structure, even in the simple case of two RM components with different intrinsic polarization angles. This is because RM Synthesis does not have an equivalent of a reduced $\chi^2$ to measure the goodness of fit as in the least square fit of  polarization position angle as a function of $\lambda^2$. As a result, the solution can  converge to an incorrect RM value. To ensure the reliability of our RMs, we have computed the reduced $\chi^2$ of the angle against $\lambda^2$ relation using the RM value obtained from RM synthesis, and we only accept RMs with $\chi^2$ $\le$ 2.  In addition, we have inspected the behavior of Stokes $Q$, $U$ and the polarized intensity as a function of  $\lambda^2$ for each EGS to ensure that the result from RM synthesis and least-square fit are in reasonable agreement. We have derived reliable RMs for 302 EGSs in the surveyed region, as listed in Table~\ref{vlatable:rm_catalogue}.

\subsection{Comparison with NVSS RMs}

\begin{figure}
\centering
\epsscale{0.5}
\plotone{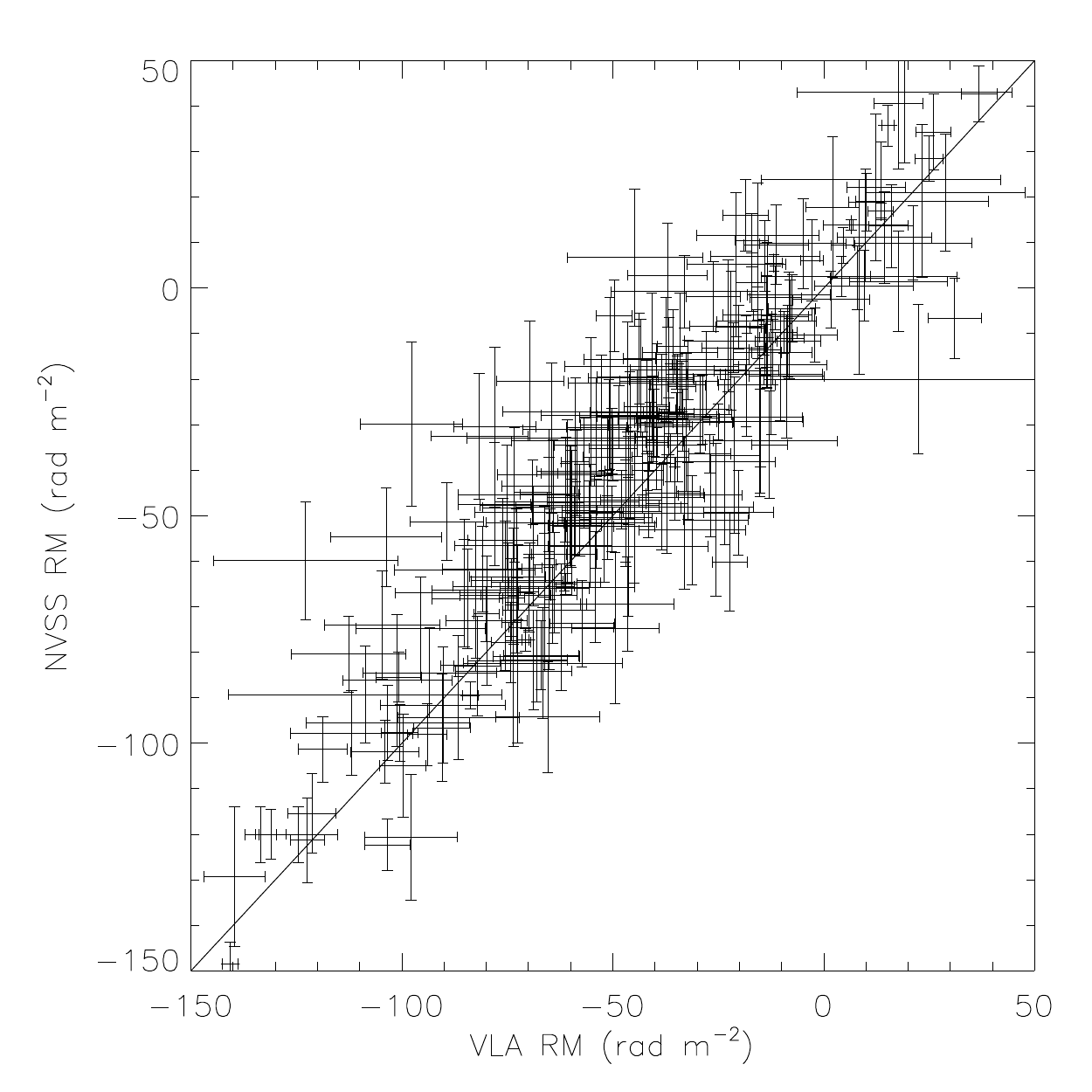}
\caption{Comparison between NVSS RMs derived by \cite{taylor2009} and our RMs in 100$^\circ$$<$$l$$<$117$^\circ$, $|b|$$<$30$^\circ$. The solid line of slope 1 indicates where NVSS RMs and our VLA RMs agree with each other.}
\label{vlafig:compare_vla_nvss}
\end{figure}

 \cite{taylor2009} have computed RMs of EGSs in the NVSS catalogue using the data from the original survey. We find matches for 244 sources of our VLA RMs in the \cite{taylor2009} catalogue. In Figure~\ref{vlafig:compare_vla_nvss}, we have plotted the RMs derived by \cite{taylor2009} against the RMs that we have derived for the same sources. The solid line of slope 1 indicates where sources should lie if the NVSS RMs and our RMs are equal. Approximately 57\% of the RMs from the two samples agree with each other within their measurement errors. The linear correlation coefficient between the two RM data sets is 0.47. However, if we compute the coefficient after discarding 3 EGS with the most extreme $|$RM$|$ in the matched data set, the linear correlation coefficient increases to 0.89, suggesting good RM agreements between the two data sets.

\subsection{RM Distribution in the Surveyed Region}

\begin{figure}
\centering
\epsscale{0.5}
\plotone{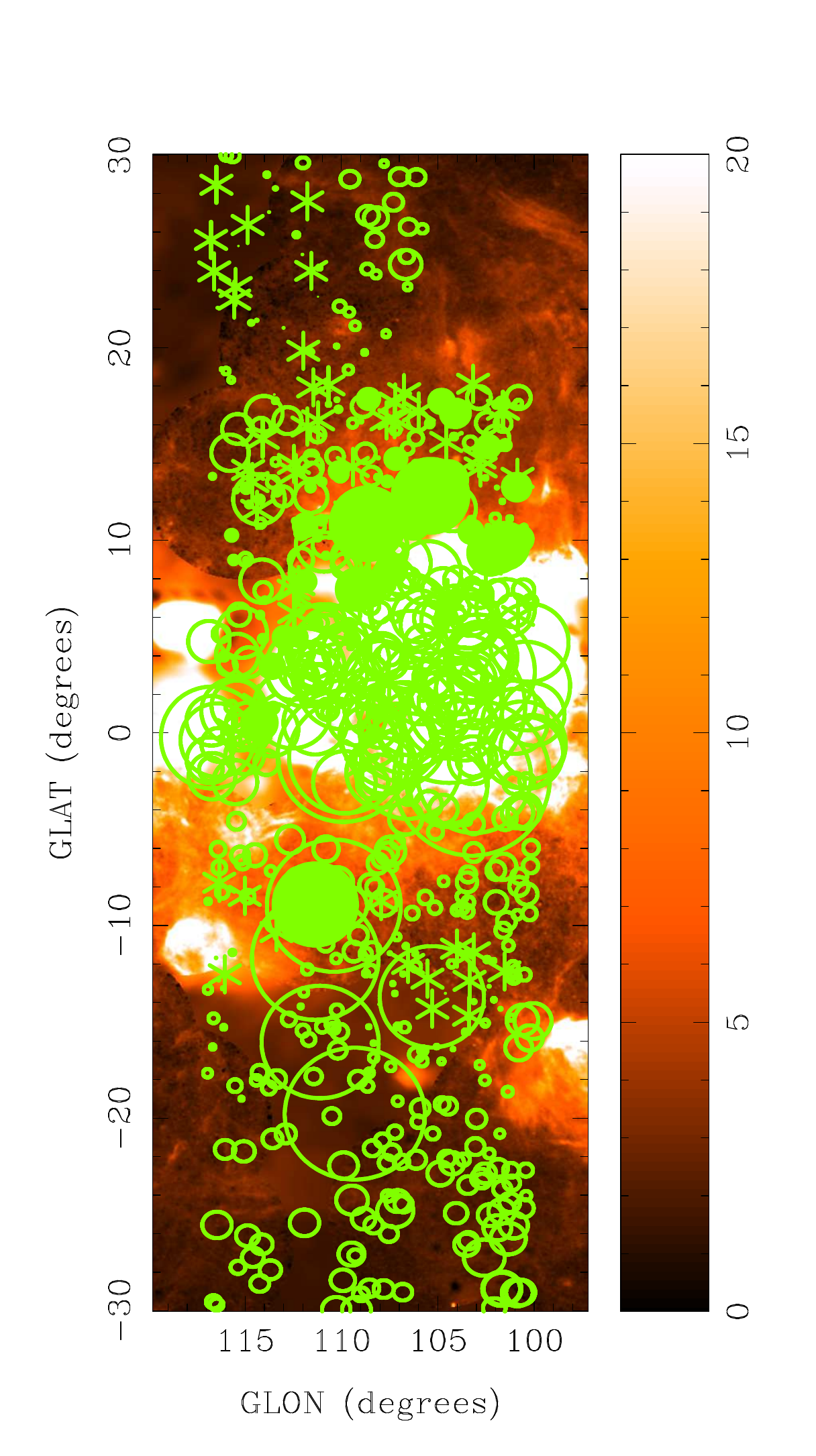}
\caption{RM distribution in the region 100$^\circ$$<$$l$$<$117$^\circ$, $|$$b$$|$$<$30$^\circ$ overlaid on an H$\alpha$ intensity map \citep{finkbeiner2003}. The color scale is in units of Rayleighs. Positive (negative) RMs are denoted by filled (open) circles with diameters proportional to $|$RM$|$. Sources with RMs consistent with zero at 1$\sigma$ are denoted by asterisks.}
\label{vlafig:rm_on_halpha}
\end{figure}

The distribution of EGS RMs in our VLA survey and in the CGPS RM survey (Brown et al. 2003, Brown et al. in prep) in the Galactic longitude range 100$^\circ$-117$^\circ$ and latitude $|$$b$$|$$<$30$^\circ$ are over-plotted on the  \cite{finkbeiner2003} all sky H$\alpha$ composite map in Figure~\ref{vlafig:rm_on_halpha}. RMs are large and negative for sight lines within 10$^\circ$ of the Galactic plane, then $|$RM$|$ slowly decreases with Galactic latitude to $|$$b$$|$$\sim$10$^\circ$. At higher latitudes ($|b|$$>$15$^\circ$), the magnitude of RM toward the northern Galactic hemisphere remains small but negative, while RMs toward the southern Galactic hemisphere slowly become more negative.

\begin{figure}
  \centering
 \subfigure{
 \epsscale{0.5}
\plotone{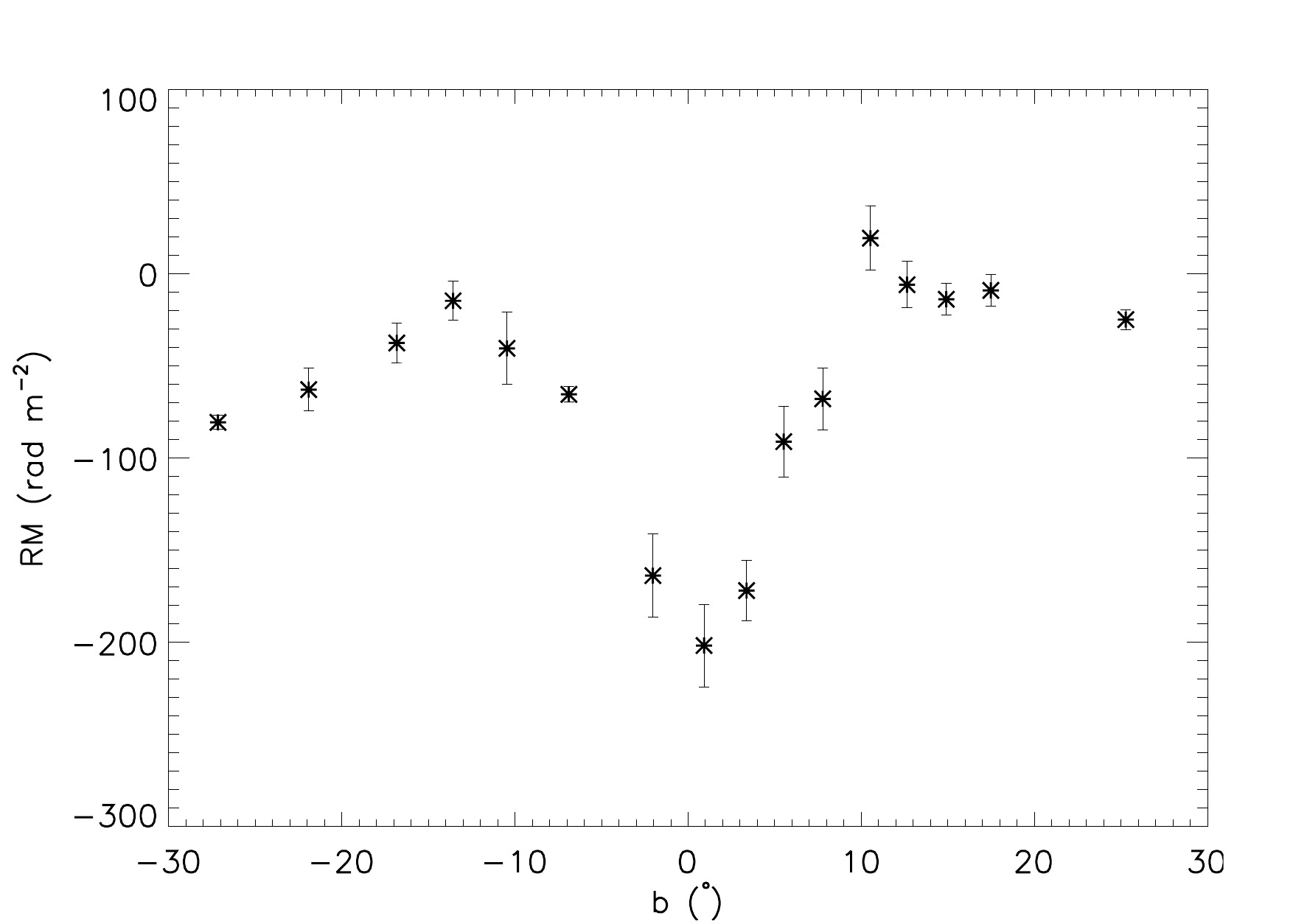}
}
\vspace{.1in}
\subfigure{
\epsscale{0.5}
\plotone{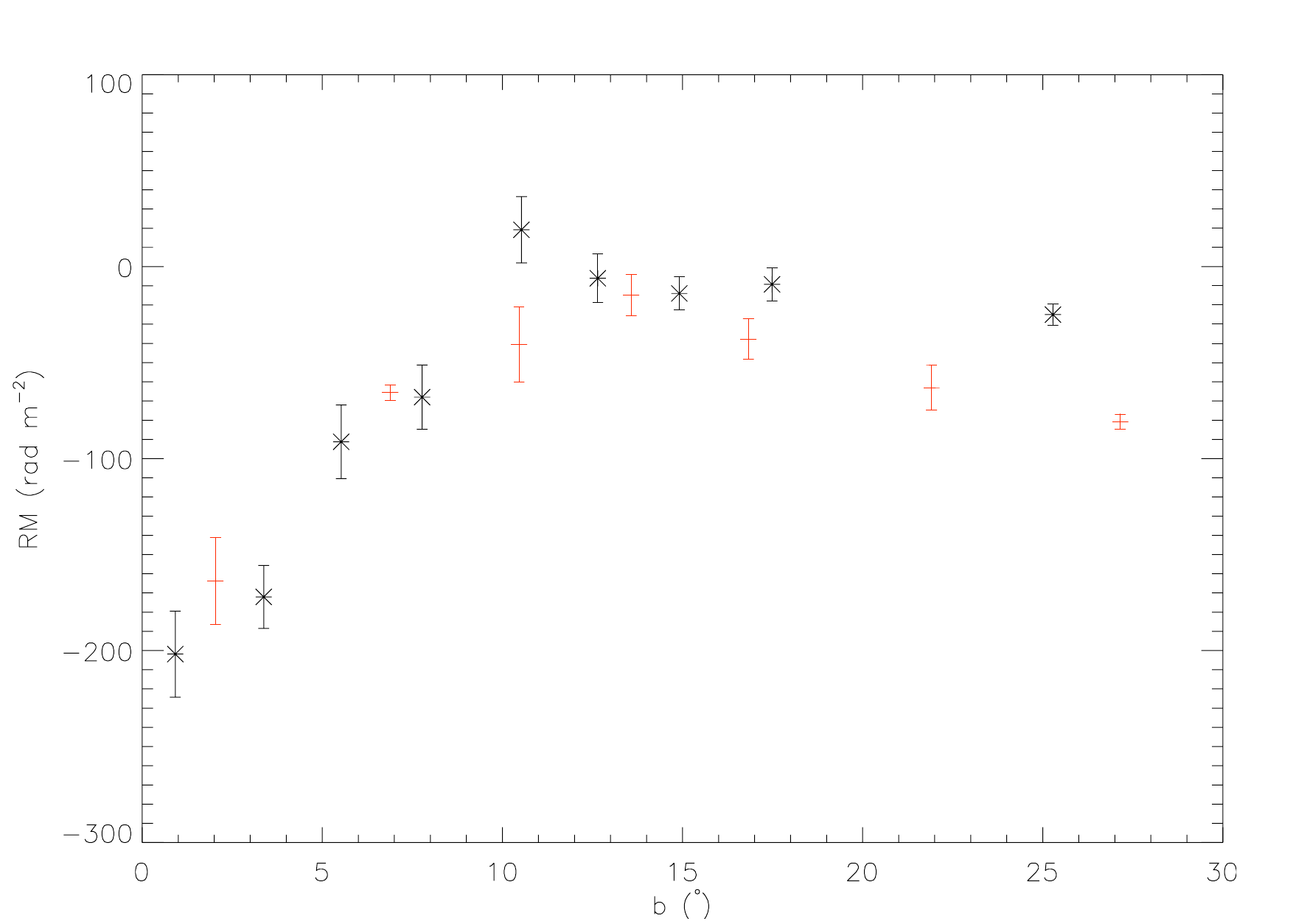}
}
\caption{Top: The variation of RM as a function of Galactic latitude for EGSs in the Galactic longitude range 100$^\circ$$<$$l$$<$117$^\circ$. The RM data are binned such that there are 40 RMs within each bin. We have plotted the median RM  and the standard error of the mean within each bin. Bottom: The variation of RM as a function of the absolute value of the Galactic latitude for EGSs in the Galactic longitude range 100$^\circ$$<$$l$$<$117$^\circ$. This plot is the same as in the top panel except that the data have been folded across $b$=0$^\circ$. Black (red) symbols represent the RM-vs-$b$ behavior above (below) the Galactic plane.}
\label{vlafig:rm_vs_b}
\end{figure}

This RM against $b$ trend is illustrated in Figure~\ref{vlafig:rm_vs_b}. In the top panel, median RMs are plotted as a function of $b$ using latitude bins containing equal number of EGS each, with error bars denoting standard error of the mean within each bin. To highlight the difference between the behavior of RM above and below the plane at $|$$b$$|$$<$15$^\circ$, we have folded the top panel about $b$=0$^\circ$, producing the figure in the bottom panel. This plot demonstrates that the RMs are symmetrically distributed close to the Galactic plane. The difference in RM for sight lines above and below the plane becomes pronounced at $|$b$|$$>$15$^\circ$.

\section{The Symmetric Large-scale Magnetic Field in the Disk Towards the Perseus Arm}
\label{vlasection:disk_symmetry} 
Magneto-ionic structures along the sight lines toward the outer Galaxy are less complex than those towards the inner Galaxy as the former only pass through one major spiral arm: the Perseus arm. For this reason and because of the high angular density of RM measurements in our new sample at 100$^\circ$$<$$l$$<$117$^\circ$, the symmetry properties of the disk field can be revealed by plotting RM  against $b$ (Figure~\ref{vlafig:rm_vs_b}) without needing to apply wavelet analysis \citep{frick2001} or averaging algorithms \citep{kronberg2011}. The signs of RM above and below the disk within $\sim$ 15$^\circ$ of the Galactic plane are the same, implying that towards the outer Galaxy, the plane-parallel disk magnetic field preserves its direction across the Galactic mid-plane (even parity), in agreement with previous works. This picture fits in well with our current understanding of the disk field origin: the dynamo mechanism predicts a symmetric disk field as it is the easiest mode to be excited \citep[see e.g.,][]{widrow2002,shukurov2004}. 

\cite{rae2010} investigated the symmetry plane of the disk field by fitting a Gaussian to the RM versus $b$ behavior using EGS RMs in the CGPS latitude extension region which mostly probe sight lines above the plane (latitude coverage $-$3.5$^\circ$$<$$b$$<$$+$18$^\circ$). They found that the RM distribution is not symmetric about $b$=0$^\circ$, but is rather shifted to the north by 1$^\circ$. \cite{rae2010} attributed this offset to a warped Galactic magneto-ionic disk. A warp in the Galactic disk towards the outer Galaxy has been detected in the stellar component \citep[for a summary, see][]{vallee2011} as well as in neutral hydrogen \citep{kalberla2009}. The distance to the warp is estimated to be at least 5 kpc from the Sun towards the outer Galaxy \citep[e.g.,][]{levine2008}. Our new VLA observations provide EGS RMs towards negative latitudes in the same longitude range as the CGPS data. This puts us at a position to re-examine results of \cite{rae2010}. 

Similar to \cite{rae2010}, we center the RM distribution about the Galactic disk by fitting a Gaussian\footnotemark[3]\footnotetext[3]{A parabolic fit to the RM distribution provides a centroid location consistent with that from a gaussian fit. However, we chose to use the Gaussian fit since it gives a lower reduced $\chi^2$.}. The Gaussian centroid fit is sensitive to the latitude range within which RMs are fitted. We use  $|$$b$$|$=10$^\circ$ as the initial boundary of the fit since it is where the RM-vs-$b$ trend changes abruptly. This is likely because RMs of EGSs at higher Galactic latitudes are dominated by the Galactic halo magnetic field which has a different vertical symmetry from the disk field. Including RMs at higher $b$ when fitting will then likely degrade the quality of the centroid fit. We vary the latitude boundary of the fit until the reduced $\chi^2$ of the Gaussian fit is minimized. We found that the distribution of rotation measure is symmetric about $b$=$-$0.13$^\circ$$\pm$0.48$^\circ$, with corresponding fit boundaries at $b$=$\pm$22.15$^\circ$. This is different from the offset of the RM symmetry axis to positive $b$ reported in \cite{rae2010}. This is likely due to the limited Galactic latitude coverage (mostly above the Galactic plane) of the CGPS data, and the fact that we have used a recently revised CGPS RM catalogue, different from that used in \cite{rae2010}. Our findings of the RM symmetry with respect to the Galactic mid-plane at low Galactic latitude is in agreement with the conclusion reached by \cite{lazio1998}, who showed, using the scattering of EGSs, that the ionized Galactic disk is not warped towards the outer Galaxy.

The fitted RM distribution has a width of 5.49$^\circ$$\pm$0.78$^\circ$: the $|$RM$|$ produced by the disk field is less than 5 rad m$^{-2}$ for $|b|$ $>$15$^\circ$. The symmetric RM behavior about the Galactic plane within $\sim$ 15$^\circ$ is consistent with the result of \cite{kronberg2011}. These authors used a smoothed all sky RM data set to show that the Milky Way's disk magnetic field stays symmetric about the mid-plane for $|$$b$$|$$<$15$^\circ$. We note that translating this latitude boundary into a physical height below which the disk field dominates requires the knowledge of where along the line of sight most of the Faraday rotation occurs. If we assume that most Faraday rotation is mainly produced by magnetic fields and electrons in the Perseus spiral arm at a distance of 2 kpc \citep{xureid2006}, then the disk magnetic field dominates within $\sim$ 540 pc of the Galactic plane.

\section{The Rotation Measure Distribution towards 100$^\circ$$<$$l$$<$117$^\circ$, 15$^\circ$$<$$|$$b$$|$$<$30$^\circ$}

In this section, we investigate the RM behavior in our surveyed longitude range at 15$^\circ$$<$$|$$b$$|$$<$30$^\circ$. We first establish the location of the Faraday rotating layer and then determine if structures in thermal electrons or in magnetic fields are responsible for producing the observed RMs. Finally, we construct a simple model of the Galactic halo magnetic field to explain the observed RM trend as a function of Galactic latitude.

\subsection{The location of the Faraday rotating layer at $|b|$$>$15$^\circ$}
\label{vlasubsection:location}
Since Faraday rotation is an integral from the EGS to the observer, the observed RM could originate anywhere along the line of sight. Therefore, it is important to determine where most of the Faraday rotation takes place. The observed RM could be of extragalactic or Galactic origin. Galaxy clusters and superclusters are magnetized and can produce enhanced RMs in directions away from the Galactic plane \citep[see e.g.,][]{kim1990,clarke2001,xu2006}. Since the mid-Galactic latitude region in our VLA survey does not coincide with extended galaxy clusters or superclusters, we rule out the possibility that the RMs we measure toward the survey area are produced in the intracluster medium.  As high velocity clouds (HVCs) can produce observable RMs \citep{mg2010}, we verified, by comparing maps of the high velocity neutral hydrogen sky \citep[e.g.,][]{wakker1997}, that no HVC lies within our surveyed region. 

The negative RMs seen towards $b$$\sim$$-$20$^\circ$ in the longitude range 100$^\circ$-117$^\circ$ in Figure~\ref{vlafig:rm_on_halpha} belong to Region A, an anomalous RM  region (60$^\circ$$<$$l$$<$140$^\circ$, -40$^\circ$$<$$b$$<$+10$^\circ$) \citep{michel1973,sk1980}. \cite{sk1980} suggested that the Sun does not reside in the region producing the negative RMs due to the lack of a positive RM region in the opposite direction. Since Region A appears to be located within the boundary of Galactic radio Loop II (the Cetus Arc), \cite{berkhuijsen1971b} suggested that the unusually large and negative RM in Region A is produced by Loop II. Loop II is thought to be the shell of a nearby ($\sim$ 100 pc) supernova remnant expanding into the magnetized interstellar medium, as by \cite{spoelstra1972} and by \cite{vallee1982,vallee1993}. In their models, a shell of thickness 10 pc, with an electron density of 1.2 cm$^{-3}$ and a magnetic field strength of $\sim$ 10 $\mu$G could produce RMs of similar magnitudes to those observed within Region A. However, one expects the thermal electrons in this shell to produce an emission measure of $\sim$ 14 pc cm$^{-6}$. This is inconsistent with WHAM observations: the mean emission measure in Region A is roughly 2  pc cm$^{-6}$, far smaller than the predicted value (since region A is far outside the Galactic plane, we expect little extinction). Furthermore, there is no enhancement in H$\alpha$ along the boundary of Region A, that might correspond to limb-brightening. One alternative is that a magneto-ionic medium which is non-H$\alpha$ emitting is responsible for producing the observed Faraday rotation. \cite{sk1980} considered the possibility that low energy relativistic electrons, which are not visible in H$\alpha$, produce the observed RM, but the estimated density of these low-energy cosmic rays was too low to explain the observed RM magnitude. \cite{sk1980} further suggested that given the dimension of Loop II, the total energy injection requires multiple supernova explosions. It is unclear how magnetic fields can stay coherent on such a large area in the sky after a few of these violent events unless they produce a single coherent superbubble. We also note that the boundary of Loop II does not lie within our survey area.

More recently, \cite{stil2011} used RMs derived from the NVSS \citep{taylor2009} to revisit the possible connection between Loop II and Region A. They argued that the RM sign-change from negative within to positive outside Region A appears to coincide with the boundary of Loop II at $l$$\sim$160$^\circ$. This was taken as evidence by \cite{stil2011} of a magnetic field reversal due to Loop II. Upon close inspection, however, we find that positive RMs outside Region A near $l$$\sim$160$^\circ$ is likely produced by ionized gas and magnetic fields in the HII region Sh2-220 (The California Nebula) and a more extended H$\alpha$ emission region around it \citep{reynolds1988,haffner1999}. The coherent magnetic field properties of Sh2-220 have been investigated in detail by \cite{harveysmith2011}. We suggest that the RM sign change is likely at least partly due to the presence of the HII region, rather than due to the absence of Loop II. 

Finally, using RMs towards pulsars in Region A, \cite{newtonmg2009} reached the conclusion that Faraday rotation along sight lines within Region A must be produced at least 0.95 kpc away from the Sun.  This once again weakens the possible connection between Region A and Loop II, the latter of which is thought to be located within $\sim$100 pc.  Therefore,  we suggest that the negative RMs that we see at mid-latitude below the Galactic plane are likely not produced by local structures in the ISM,  but rather by larger scale structures located at least 1 kpc from the Sun. 

\begin{figure}
\centering
\epsscale{0.7}
\plotone{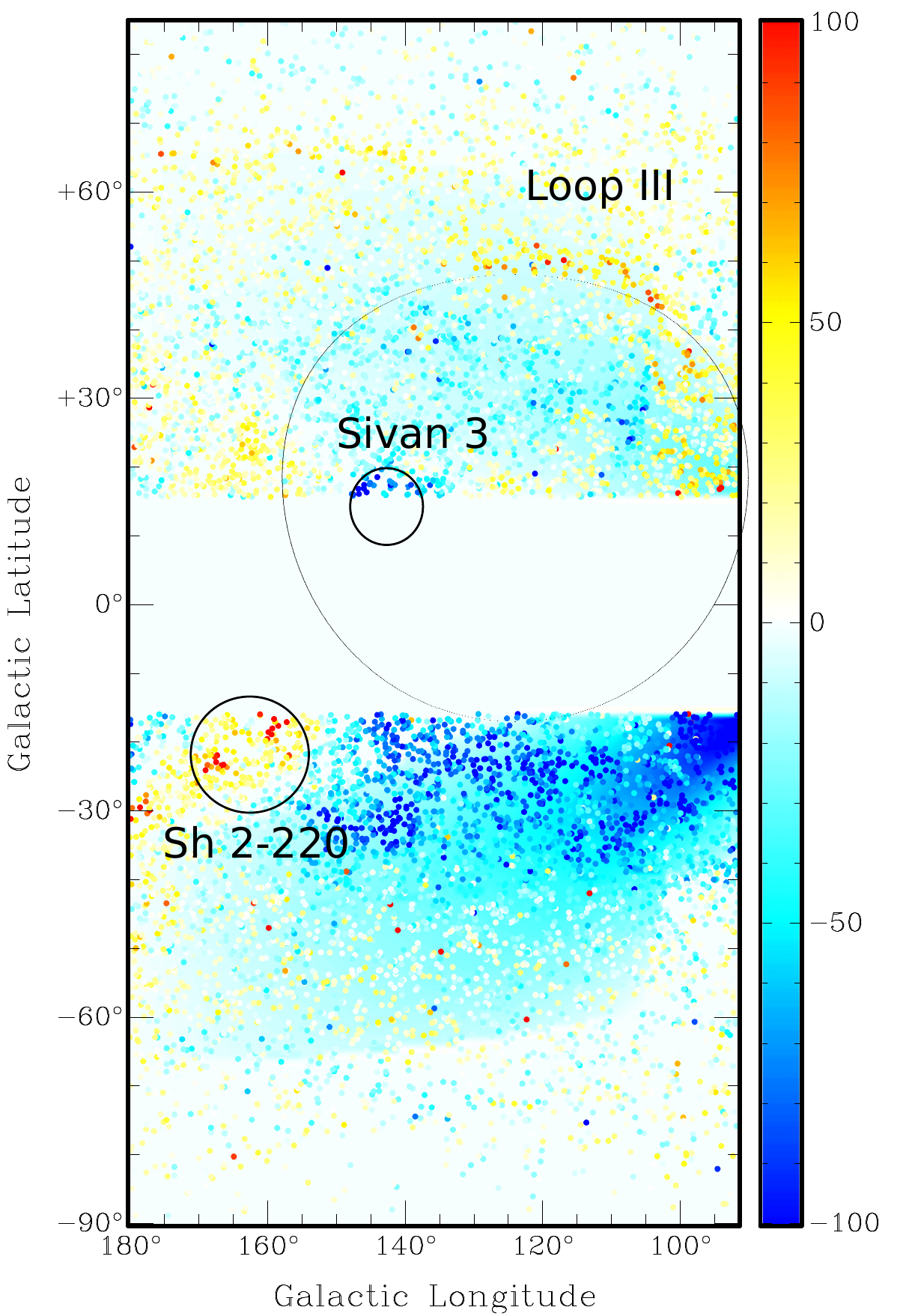}
\caption{The RM predicted from our simple halo field model towards the entire second Galactic quadrant viewed as a 2D distribution: the color scheme is chosen such that red (blue) represents positive (negative) RMs, while white represents RMs close to 0 rad m$^{-2}$. RMs of extragalactic sources from \cite{taylor2009} are over-plotted as filled circles on the same color scheme. Three large regions where there is discrepancy between the model fit and the RM data are labeled: Loop III and HII regions Sivan 3 and Sh 2-220. The Milky Way disk magnetic field dominates within 15$^\circ$ of the Galactic plane and therefore the region is blanked.}
\label{vlafig:rm_prediction_q2}
\end{figure}

Our surveyed region above the Galactic plane lies within the boundary of Radio Loop III \citep{berkhuijsen1971b}. Although RMs at the low Galactic longitude edge of Loop III appear to be enhanced (see~Figure \ref{vlafig:rm_prediction_q2}), the RM distribution within Loop III itself does not appear to be anomalous\footnotemark[4]\footnotetext[4]{We note that the variance of RM within Loop III was reported to be enhanced by \cite{stil2011}.}.

\subsection{Enhancement in Electron Density or Magnetic Field?}
\label{vlasubsection:enhancement}
Since RM is sensitive to magnetic field weighted by thermal electron density, an increase in EGS RM could reflect either an increase in the average line-of-sight magnetic field strength, in the thermal electron density or in both quantities. If thermal electrons are responsible for producing the observed RM as \cite{sk1980} argued, and the same thermal electrons also emit in H$\alpha$, then one can distinguish between RM due to an increase in line-of-sight magnetic field strength from that due to an increase in electron density by inspecting the distribution of H$\alpha$ emission along the sight lines of interest. 

We found that for both local ($-$25 km s$^{-1}$$<$v$_{LSR}$$<$$+$25 km s$^{-1}$)  and Perseus arm  ($-$60 km s$^{-1}$$<$v$_{LSR}$ $<$ $-$40 km s$^{-1}$) ionized gas, the emission measure is distributed symmetrically about the Galactic plane -- there is comparable amount of thermal electrons at +20$^\circ$$<$$b$$<$+30$^\circ$ as that towards $-$30$^\circ$$<$$b$$<$$-$20$^\circ$. Therefore,  the structure in RM is unlikely to be due to enhancements in thermal electrons. Instead, it is likely that the observed RM in the surveyed region reflects changes in magnetic field structures. \cite{sk1980} also pointed out that if the increase in RM was primarily due to an increased electron density, then the electron density would be a factor of 50 larger towards the southern Galactic hemisphere than towards the northern Galactic hemisphere (this is roughly the ratio of $|$RM$|$ at $b$=$-$20$^\circ$ to  that at $b$=$+$20$^\circ$). If the pulsars were beyond the Faraday rotating region, their DMs would reflect this increase in electron densities. If that is the case, then pulsars at $b$$>$$+$20$^\circ$ should on average have much smaller DMs than those at $b$$<$$-$20$^\circ$. Both \cite{sk1980} and \cite{newtonmg2009} did not find evidence for such a pulsar DM distribution, supporting our claim that RMs in our VLA survey likely reflect structures in magnetic fields rather than electron densities. 


\subsection{Prediction of RM from Existing Halo Magnetic Field Models}
\label{vlasubsection:halo_models}

In Sections \ref{vlasubsection:location} and \ref{vlasubsection:enhancement}, we have established that RMs towards mid-latitudes in the longitude range 100$^\circ$-117$^\circ$ are of Galactic origin and are likely due to large-scale structures. We also reach the conclusion that RMs at mid-Galactic latitudes likely reflect structures in magnetic fields rather than electron densities. In this section, using expressions for the halo magnetic field in the literature, we show that no existing models are successful in reproducing the observed EGS RMs for sight lines $|$$b$$|$$>$15$^\circ$ in the longitude range 100$^\circ$$<$$l$$<$117$^\circ$. It is evident that additional modeling is required to properly characterize the Galactic halo magnetic field. 

The median RM for $b$$>$+15$^\circ$ is $-$15$\pm$4 rad m$^{-2}$ while the median RM for $b$$<$$-$15$^\circ$ is $-$62$\pm$5 rad m$^{-2}$. Since the median RM is negative both above and below the plane at mid-Galactic latitudes ($|$$b$$|$$>$15$^\circ$), we suggest that the large-scale halo magnetic field parallel to the plane toward the outer Galaxy does not reverse direction across the mid-plane. This is consistent with recent findings of \cite{pavel2012}: these authors used Near IR polarimetry of starlight towards $l$ $=$ 150$^\circ$ between $-$75$^\circ$ $<$b$<$ $+$10$^\circ$ and simulations to rule out magnetic fields of odd parity towards the outer Galaxy. Since we did not find an asymmetric electron density distribution about the Galactic mid-plane at $|$b$|$$>$20$^\circ$  in Section~\ref{vlasubsection:enhancement}, the different RM magnitudes or orientations above and below the plane are likely due to different halo magnetic field strengths in the northern/southern hemispheres. 

Expressions for the halo magnetic field in the literature often contain a magnetic field reversal across the Galactic plane. \cite{han1997} and \cite{han2002} claim, based on the anti-symmetric sign of EGS RMs at high Galactic latitudes towards the inner Galaxy, that the Milky Way's halo magnetic field is dipole-like (for which the horizontal component reverses direction across mid-plane, but the vertical component does not). This claim is challenged by \cite{mao2010}, who demonstrated that the magnetic field geometry at the location of the Sun cannot be a pure dipole or a quadrupole. Moreover, \cite{wolleben2010} used RMs determined from diffuse synchrotron polarized emission to show that a nearby magnetized bubble could produce part of the anti-symmetric EGS RM pattern that \cite{han1997} interpreted as a signature of a global dipole-like field. Although models in the literature that predict a reversal in magnetic field direction across the plane towards Galactic quadrant 2 are unlikely to be physical in light these more recent works, we nonetheless plot the predictions from these models in this section. 

Existing halo magnetic field models consist of two general types: toroidal fields \citep[e.g.,][]{sun2008} or exponential plane-parallel halo fields that fall off with Galacto-centric radius and distance from the mid-plane. Parameters of the halo field models can be divided into two different groups as well: values that can qualitatively reproduce the halo field observables (RMs of EGSs, diffuse synchrotron emission and starlight polarization)\citep[e.g.,][]{sun2008}, or values that are obtained by finding a best fit to the halo field observables, minimizing the $\chi^2$ \citep[e.g.,][]{jansson2009}. 

In the following, we list all the halo field expressions in the literature. We define a parameter $S$ that captures the dipolar nature of the halo field: its value is $-$1 for $b$$>$0$^\circ$ and +1 for $b$$<$0$^\circ$. The sign is chosen such that the halo field points in the same direction (clockwise) as the disk field below the plane and reverses such that it is (counterclockwise) above the plane in Galactic quadrant 2 when viewed from the north Galactic pole. The halo field expressions are given in a cylindrical Galacto-centric coordinate system, for which components in the radial ($r$), azimuth ($\phi$) and vertical ($z$) directions are listed separately. We also provide the corresponding model parameters and note  whether these parameters are found qualitatively by eye or by a quantitative fit to halo field observables.

The \cite{jansson2009} halo field ($E2_{N0,2009}$) is given by:
\begin{align}
{\rm For}~r<r_c  \notag \\
 B_r &=  S B^H_0 sin(p_{inner}), \notag \\
B_\phi & = -S B^H_0 cos(p_{inner}), \notag \\
B_z &  = 0 \notag. \\
{\rm For} ~r\ge r_c  \notag \\
B_r & =  B^H_0 sin(p_{outer}), \notag\\
B_\phi & = -B^H_0 cos(p_{outer}), \notag \\
B_z & = 0.
\end{align}
The best fit parameters are: $r_c$=8.72 kpc, the radius beyond which the horizontal field reverses across the mid-plane; $B^H_0$ = 2.3 $\mu$G, the strength of the horizontal field; and $p_{inner}$=$-$2$^\circ$ ($p_{outer}$ = $-$30$^\circ$), the pitch angle for $r<r_c$ ($r>r_c$).

\begin{figure}
\centering
\epsscale{0.6}
\plotone{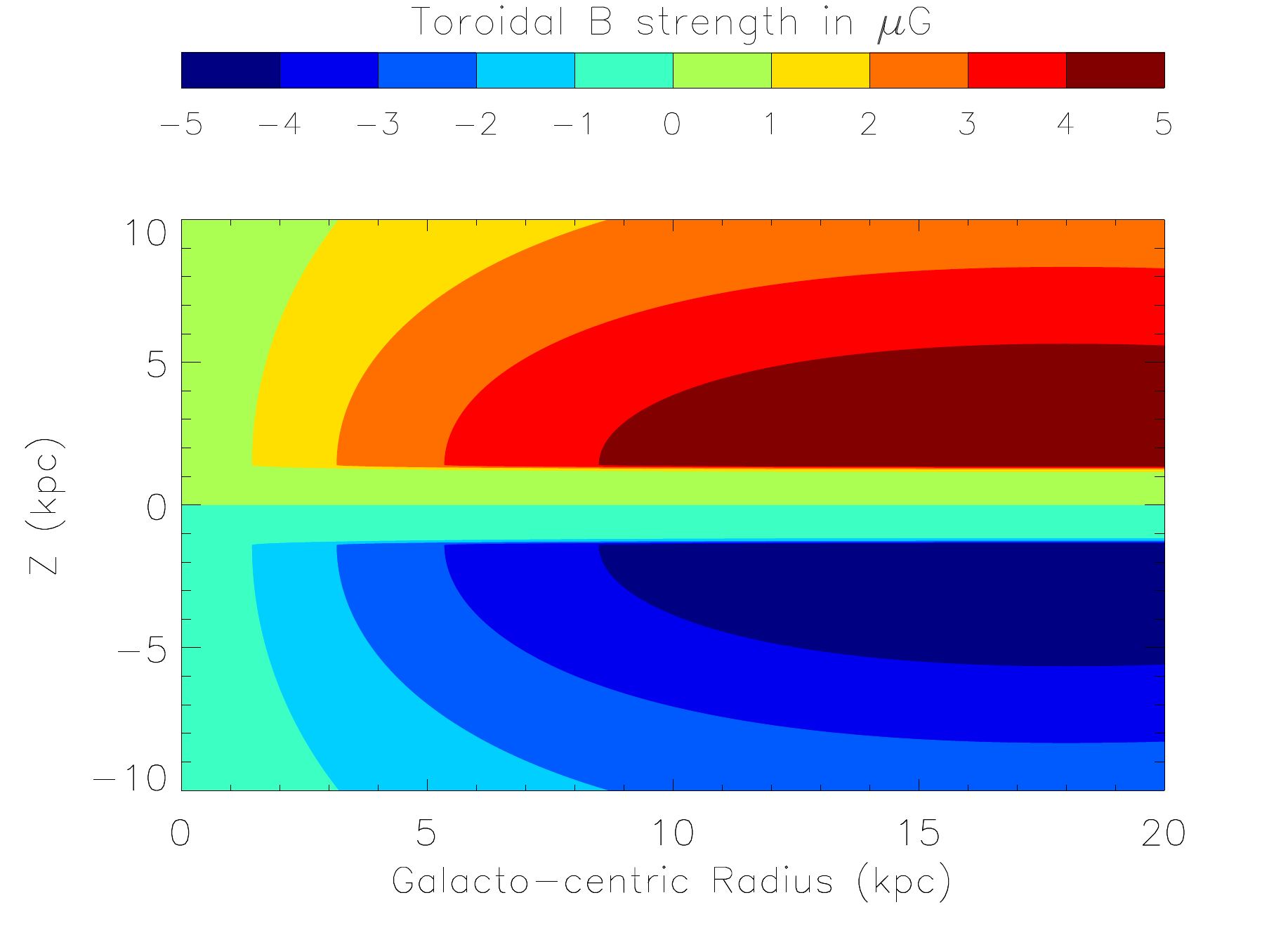}
\vspace{.05in}
\epsscale{0.6}
\plotone{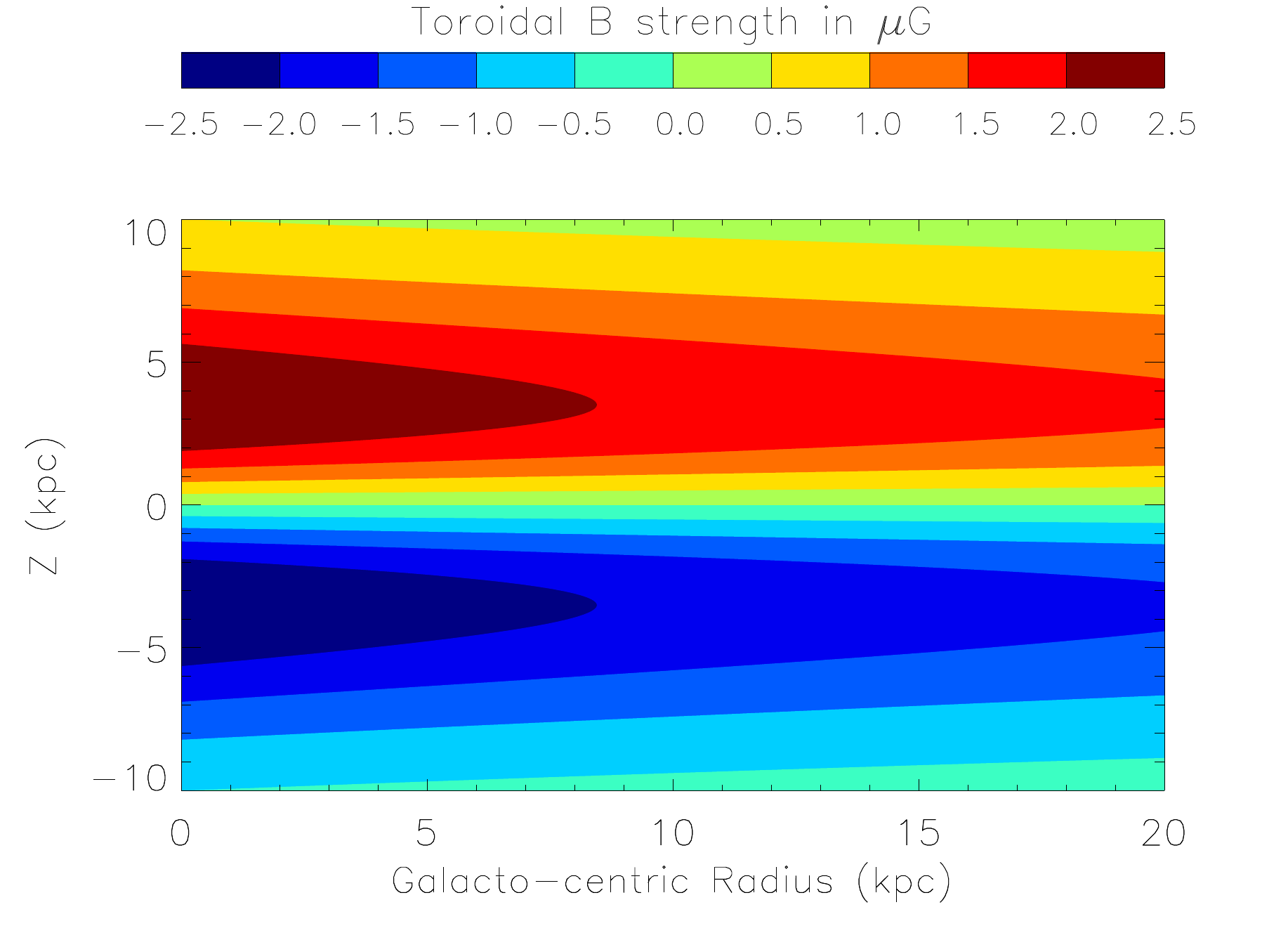}
\caption{Toroidal magnetic field strength in a vertical cross-section of the Milky Way using the \cite{jansson2009} best fit double torus model (top panel) and the \cite{rg2010} model (bottom panel).}
\label{vlafig:visualizeb}
\end{figure}

A variation of the \cite{ps2003} double torus magnetic field of the following form has been heavily used in the literature:
\begin{align}
B_r &=  0, \notag \\
B_\phi &= -S B^H_0 \frac{1}{1+\big(\frac{|z|-z^H_0}{z^H_1}\big)^2}\frac{r}{r^H_0} exp(-\frac{r-r^H_0}{r^H_0}), \notag \\
B_z & = 0. 
\end{align}
The values of $B^H_0$, $r^H_0$,$z^H_0$ and $z^H_1$ used by \cite{sun2008},  \cite{jansson2009}, \cite{sun2010} and \cite{pshirkov2011} are summarized in Table~\ref{vlatable:ps_field_form}. The nature of these parameters (qualitative or fit) are listed in the last column of Table~\ref{vlatable:ps_field_form}. To illustrate the magnetic field strength of the double torus model, we have plotted the azimuthal magnetic field strength using the best fit parameters of \cite{jansson2009} in a vertical cross-section of the Milky Way in the top panel of Figure~\ref{vlafig:visualizeb}.

Finally, \cite{rg2010} used a bi-toroidal halo field:
\begin{eqnarray}
&& B_r =  0, \nonumber \\
&& B_\phi = \frac{3r_1+24}{r1+r}arctan(\frac{z}{\sigma_1}) exp\big(-\frac{z^2}{2\sigma_2^2}\big), \nonumber \\
&& B_z = 0.2. \nonumber \\
\end{eqnarray}
The corresponding azimuthal magnetic field strength in a vertical cross-section of the Milky Way is shown in the bottom panel of Figure~\ref{vlafig:visualizeb}. The best fit values are $r_1$ $>$ 33.8 kpc, $\sigma_1$ = 2.9 kpc and $\sigma_2$ $>$ 4.7 kpc.

\begin{figure}
\centering
\epsscale{0.8}
\plotone{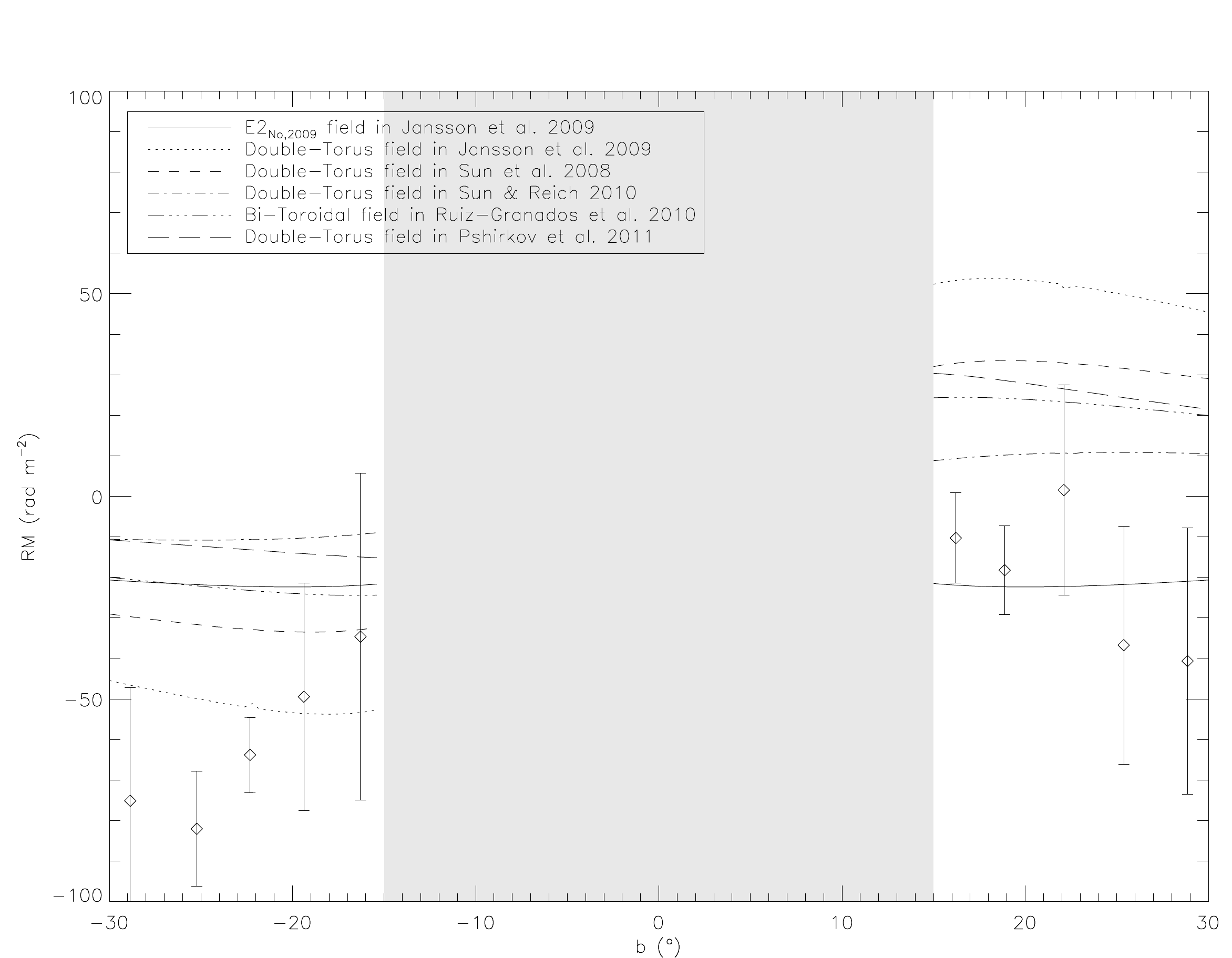}
\caption{The trend of RM as a function of Galactic latitude towards $l$=108.5$^\circ$. RM predictions from  6 different halo field models in the literature have been plotted. We assume that the halo magnetic field dominates at a height of 540 pc, which provides an upper limit for $|$RM$|$. (Models with larger transition heights have fewer thermal electrons and hence smaller $|$RM$|$). The VLA and CGPS RMs are binned every 3$^\circ$: diamonds represent the median RM within each bin and the error bars denote the weighted standard deviation within each bin. For $|$$b$$|$$<$15$^\circ$ (shaded region), the disk field dominates and therefore is not shown for clarity.}
\label{vlafig:plot_all_halo_model}
\end{figure}

In computing expected RMs from various halo field models, we assume that the Sun is located  $R_{sun}$= 8.4 kpc  from the Galactic center \citep{reid2009}. We further assume a disk-field-to-halo-field transition height\footnotemark[5]\footnotetext[5]{Below this height, the disk magnetic field dominates and the coherent halo field is assumed to be zero. Above this height, the halo magnetic field dominates and the coherent disk field is assumed to be zero.} of  $\sim$ 540 pc (as inferred from the RM-vs-$b$ trend presented in Section~\ref{vlasection:disk_symmetry}). For the Faraday rotating electrons, we use the thermal electron thick disk model of \cite{gaensler2008}, in which the WIM  is modeled as an exponential disk of scale height $h_{\rm WIM}$=1.83 kpc with a mid-plane density near the Sun of $n_{e,0}$=0.014 cm$^{-3}$. We adopt the radial dependence of the thick disk used by \cite{cl2002} with a truncation at Galacto-centric radius A$_1$= 20 kpc. The exact electron density distribution used is
\begin{equation}
n_e = n_{e,0} \frac{cos(\frac{\pi r}{2A_1})}{cos(\frac{\pi R_{sun}}{2A_1})} exp(\frac{-|z|}{h_{\rm WIM}}).
\end{equation}
\cite{cl2002} adopt a vertical distribution in the form of sech$^2$(z/h$_{\rm WIM}$) rather than the exponential in the above expression. We verify that the two z dependence produce RM against Galactic latitude trends that are very similar. Since we are focusing on fitting mid-latitude RMs, we do not seek to fit to EGS RMs at $|$$b$$|$$\le$15$^\circ$. We numerically integrate the thermal electron density weighted magnetic fields toward $l$=108.5$^\circ$, the median longitude of our surveyed area. The predicted RM-vs-$b$ trends are shown in Figure~\ref{vlafig:plot_all_halo_model}. We note that a low disk-to-halo transition height is strongly favored simply because there are more thermal electrons closer to the mid-plane, which produce larger RMs given the same magnetic field distribution. If one adopts larger transition heights, the predicted $|$RM$|$ will be smaller than that shown in Figure~\ref{vlafig:plot_all_halo_model} due to the decrease in thermal electron density at larger distances from the Galactic plane.

It is clear from Figure~\ref{vlafig:plot_all_halo_model}  that existing Milky Way halo magnetic field models do poorly in reproducing the observed RM-vs-$b$ trend for both positive and negative latitudes in the surveyed region. None of the models can simultaneously reproduce the symmetry property of the observed RM and the magnitude of the RMs. The disagreement between observed RMs and model predictions in this $l$ and $b$ range have been noted by several authors. Both \cite{sun2008} and  \cite{pshirkov2011} have attributed the discrepancy to radio continuum loops  and Region A, but we have shown in Section~\ref{vlasubsection:location} that these are unlikely to be the cause.

\subsection{A Model of the Milky Way Halo Field towards 100$^\circ$$<$$l$$<$117$^\circ$}
\label{vlasubsection:B_toy_model}

We first attempt to vary parameters of the existing halo field models to match our observed RMs. Unfortunately, adjusting the parameters to match the observed RM magnitudes produces excess RM towards the Galactic poles, for which the observed values are zero towards the north and +6 rad m$^{-2}$ towards the south \citep{taylor2009,mao2010}. Instead of fitting the observed RM with a global exponential field or a double toroid above and below the plane, we choose to approach the problem by constructing a very simple model of a part of the Milky Way's magnetic halo to reproduce the observed RM trend.

Since we reside in the Milky Way, drawing conclusions on the Galactic halo magnetic field configuration based on EGS RMs is challenging without a bird's eye view. Observations of external galaxies have revealed X-shaped magnetic fields in edge-on galaxies \citep[e.g.,][]{heesen2009} and axisymmetric spiral with an out-of-plane quadrupole extension in face-on spiral galaxies \citep[e.g.,][]{braun2010}. None of these external galaxies possess magnetic field structure similar to the Milky Way halo field models proposed in the literature: a pure toroidal field or a simple exponential field. Therefore, current  Galactic halo field models might be unrealistic.

On the theoretical front, the popular double-torus model might be physically too simplistic as its pitch angle is zero everywhere in the halo. It is thought that a pure circular magnetic field cannot be maintained  by any velocity field against turbulent magnetic diffusion \citep{beck1996}. Furthermore, if the halo magnetic fields were of dynamo origin, then spiral magnetic field lines would form naturally, since the dynamo mechanism relies on differential rotation to transform radial fields into azimuthal fields. Motivated by both observation and theory, we seek a new halo magnetic field model to explain the RMs in our observed  region.

We assume that there is a constant, coherent azimuthal magnetic field $B^H_0$ parallel to the Galactic plane in Galacto-centric radius range $r_{\rm inner}<r<r_{\rm outer}$ kpc and at a height  $z_{\rm lower}>|z|>z_{\rm upper}$ kpc from the Galactic mid-plane. We assume the halo magnetic field strength outside this specified region is zero:
\begin{align}
{\rm For}~r_{\rm inner}<r<r_{\rm outer},  &&\notag \\
z_{\rm lower}<\left | z\right |<z_{\rm upper}: &&\notag \\
 B_r &=  0, \notag \\
  B_\phi & = -B^H_0, \notag \\
B_z &  = 0; \notag \\
{\rm otherwise: } && \notag \\
B_r & =  B_\phi  = B_z = 0.
\end{align}
We allow different values of $B^H_0$ above and below the mid-plane and assume that the Faraday rotating medium is the WIM thick disk \citep{gaensler2008}.

\begin{figure}
\centering
\epsscale{0.8}
\plotone{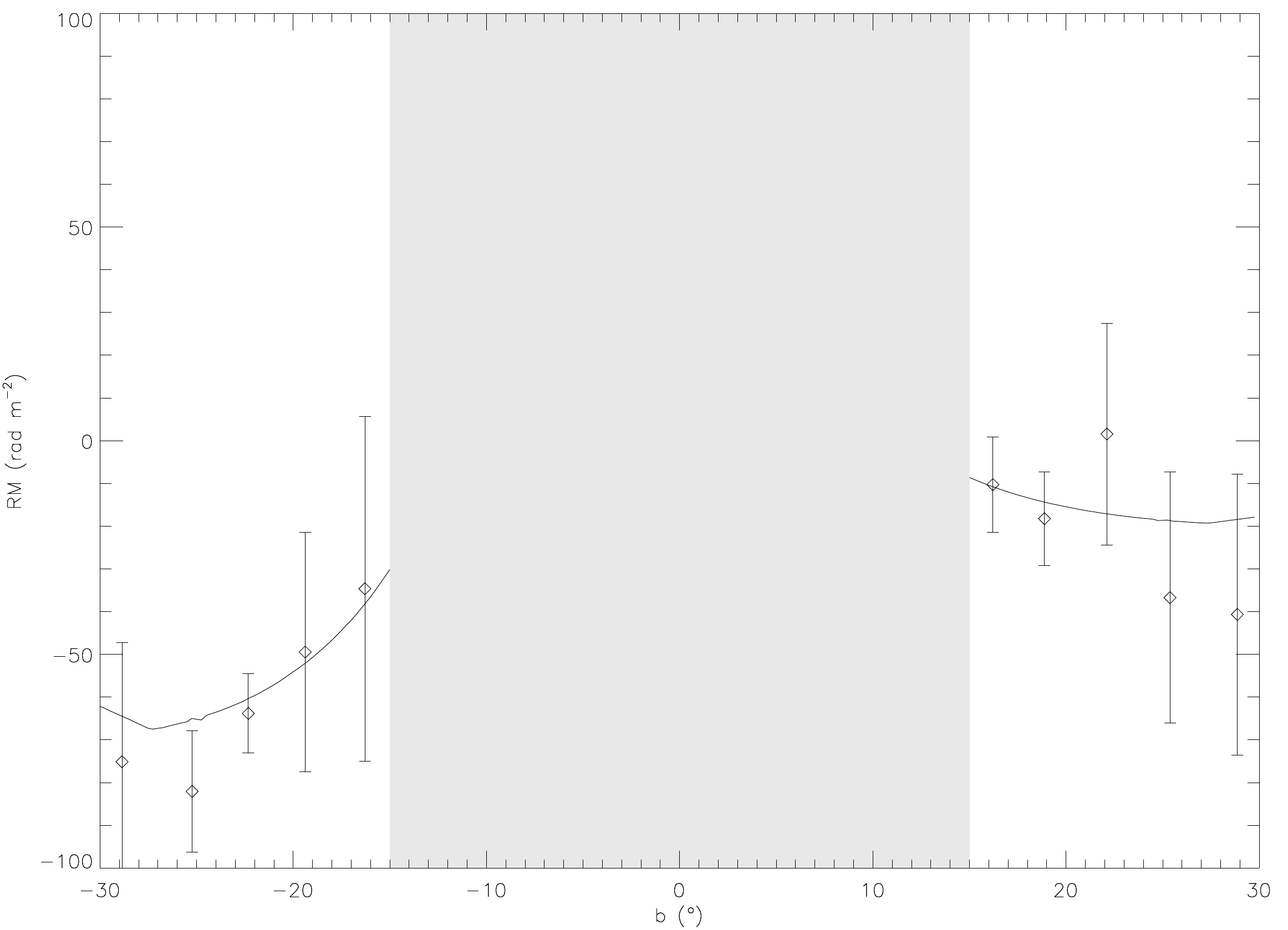}
\caption{The trend of RM as a function of Galactic latitude towards $l$=108.5$^\circ$. The RM prediction from our halo field model in Section~\ref{vlasubsection:B_toy_model} (solid line) is over-plotted on the binned VLA and CGPS RMs denoted by diamonds. The error bars are the weighted standard deviation within each bin. For $|$$b$$|$$<$15$^\circ$ (shaded region), the disk field dominates.}
\label{vlafig:B_toy_model_rm_fit}
\end{figure}

\begin{figure}
\centering{
\epsscale{0.4}
\plotone{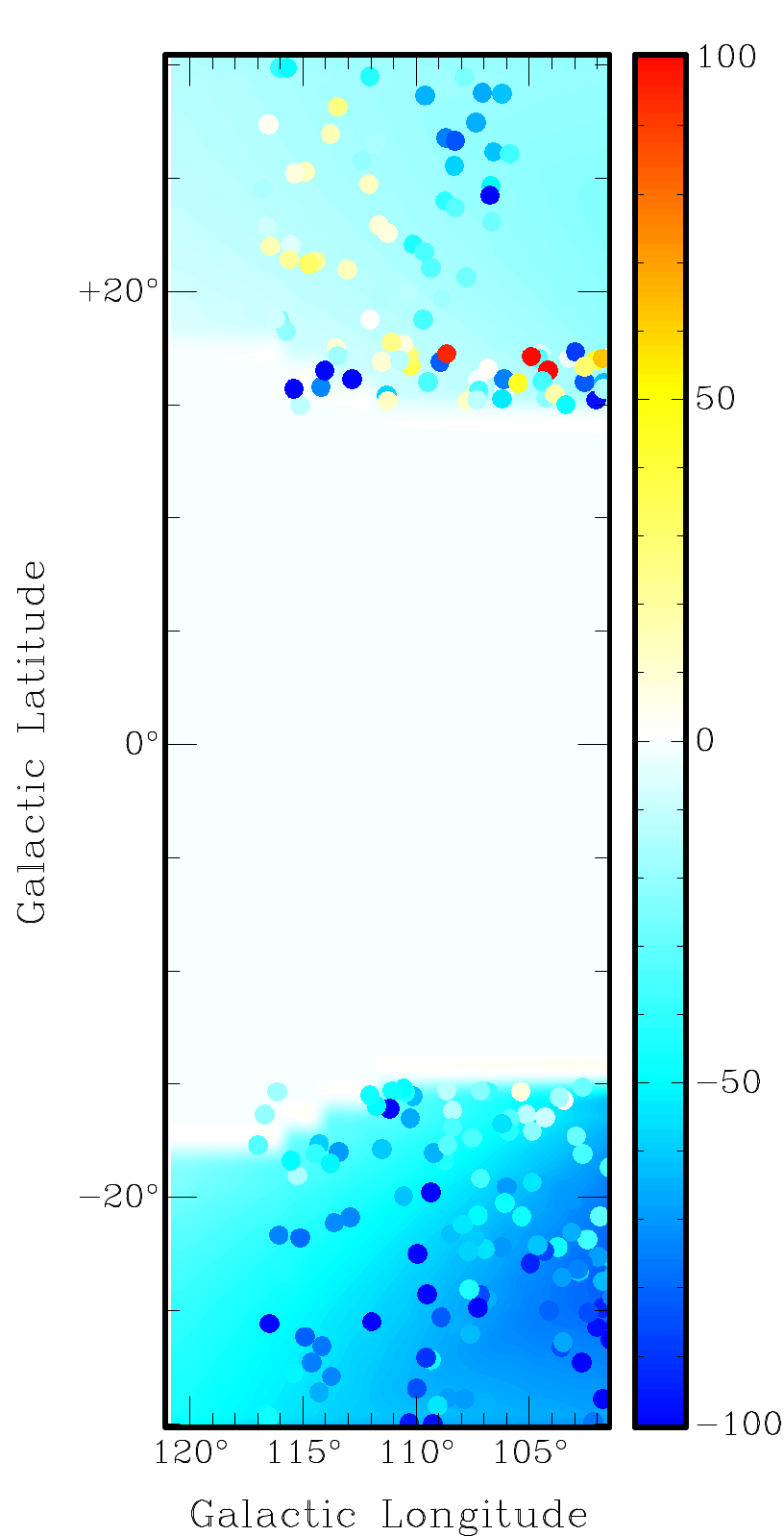}
\epsscale{0.4}
\plotone{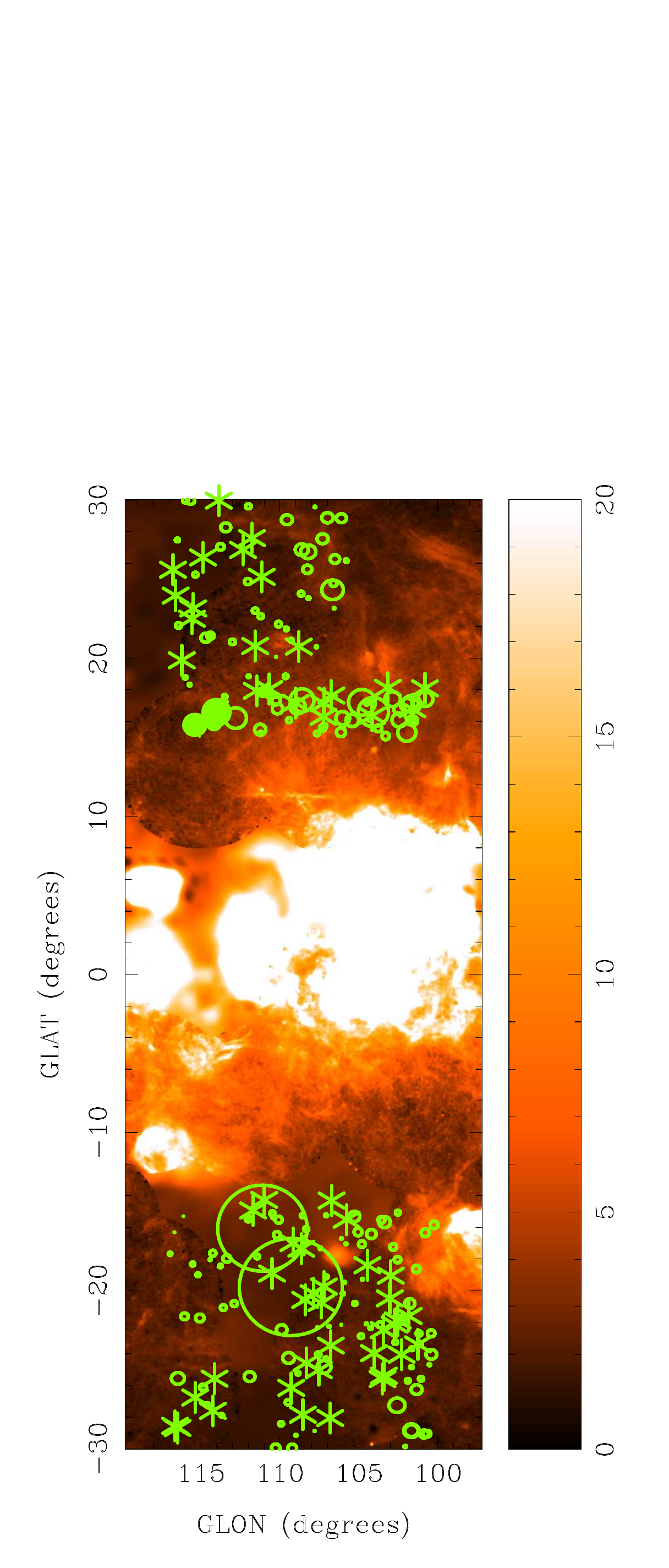}
}
\caption{Left Panel: The RM predicted from our simple halo field model shown as a 2D distribution: the color scheme is chosen such that red (blue) represents positive (negative) RMs, while white represents RMs close to 0 rad m$^{-2}$. Measured RMs of EGSs are overplotted as filled circles on the same color scale. Color differences between filled circles and the background shows the discrepancy between the model prediction and the observed RMs. We note that the uncertainties associated with individual EGS RMs are not represented in this image. The color difference between the observed and predicted RM in latitude range +15$^\circ$$<$$b$$<$+20$^\circ$ is likely due to the larger measurement errors of the CGPS RMs compared to the VLA RMs. Right Panel: The distribution of the residual RM (the difference between the predicted RMs from our halo field model and the measured RMs) overlaid on an H$\alpha$ intensity map \citep{finkbeiner2003}. The color scale is in units of Rayleighs. Positive (negative) residual RMs are denoted by filled (open) circles with diameters proportional to the amount of residual. Asterisks represent sight lines with residual RMs consistent with zero at 1$\sigma$.}
\label{vlafig:B_toy_model_rm_fit_2D}
\end{figure}

We found by performing a search in parameter space that a clockwise toroidal halo field with $B^H_0$ = 2 $\mu$G  above the plane and $B^H_0$ = 7 $\mu$G below the plane,  $r_{\rm inner}$ = 8.8 kpc, $r_{\rm outer}$ = 10.3 kpc, $z_{\rm lower}$=0.8 kpc and $z_{\rm upper}$ = 2 kpc can successfully reproduce the RM-against-$b$ trend in our observed area for $|$$b$$|$$>$15$^\circ$. This is demonstrated in Figure~\ref{vlafig:B_toy_model_rm_fit}, where the predicted RM variation as a function of latitude towards  $l$=108.5$^\circ$ (the average Galactic longitude of the observed region)  is plotted. The shape of the model-predicted RM closely matches that of the data. The 2D spatial distribution of the predicted RMs in the surveyed area is plotted in the left panel of Figure~\ref{vlafig:B_toy_model_rm_fit_2D}, where EGS RMs at $|b|$$>$15$^\circ$ are overlaid as circles with the same color scale. The color of filled circles and that of the background (prediction from our simple model) are in general agreement. The right panel of Figure~\ref{vlafig:B_toy_model_rm_fit_2D} shows the distribution of the residual RMs after subtracting the model prediction from the observed RMs. The residual RMs are small and mostly consistent with zero. We have compared the reduce  $\chi^2$ of our simple model to all of the models presented in Section~\ref{vlasubsection:halo_models} at $|b|$$>$15$^\circ$, in the Galactic longitude range 100$^\circ$$<$$l$$<$117$^\circ$. Our simple model is superior to existing halo field models at larger than 99.9 \% confidence level in the region of interest\footnotemark[6]\footnotetext[6]{We take notice that since the existing halo field models are global fits to the entire high latitude RM sky, rather than a local fit to the RM distribution in the region 100$^\circ$$<$$l$$<$117$^\circ$, 15$^\circ$$<$$|b|$$<$30$^\circ$. Therefore, the quality of fit of these global models could be degraded within our specific region of interest when attempting to reproduce the RM distribution elsewhere.}

The modeled coherent magnetic field exists at a distance between 0.8 and 2 kpc from the Galactic disk, which is well above the $z$ height where the disk field dominates ($\sim$ 540 pc, Section~\ref{vlasection:disk_symmetry}). Hydrostatic equilibrium in the Galactic halo predicts a magnetic field strength of $\sim$ 3-4 $\mu$G at a distance 1-2 kpc from the mid-plane \citep{kalberla1998,fletcher2001,cox2005}. The best-fit magnetic field strengths towards the northern and the southern Galactic hemispheres of 2 and 7 $\mu$G are in rough agreement with hydrostatic equilibrium. The need for a strong (7$\mu$G) field could be a consequence of the simplicity of our model:  we have set the coherent magnetic field strength to be zero external to Galactocentric radii in the range 8.8-10.3 kpc. Although our best-fit value for the magnetic field strength in the southern halo is comparable to that derived in the double-torus model used by \cite{sun2008}, our model is unlikely to over-predict the corresponding synchrotron emission as did the \cite{sun2008} model. This is because our 7$\mu$G halo magnetic field has a much lower  volume filling factor than the \cite{sun2008} double-torus field model. In our model, the distance to the region where Faraday rotation occurs is approximately 1 kpc, similar to the limit implied from pulsar RMs (see Section~\ref{vlasubsection:location}). Therefore, the best-fit parameters of our model appear to be consistent with other observables. 

\begin{figure}
\centering
\epsscale{0.8}
\plotone{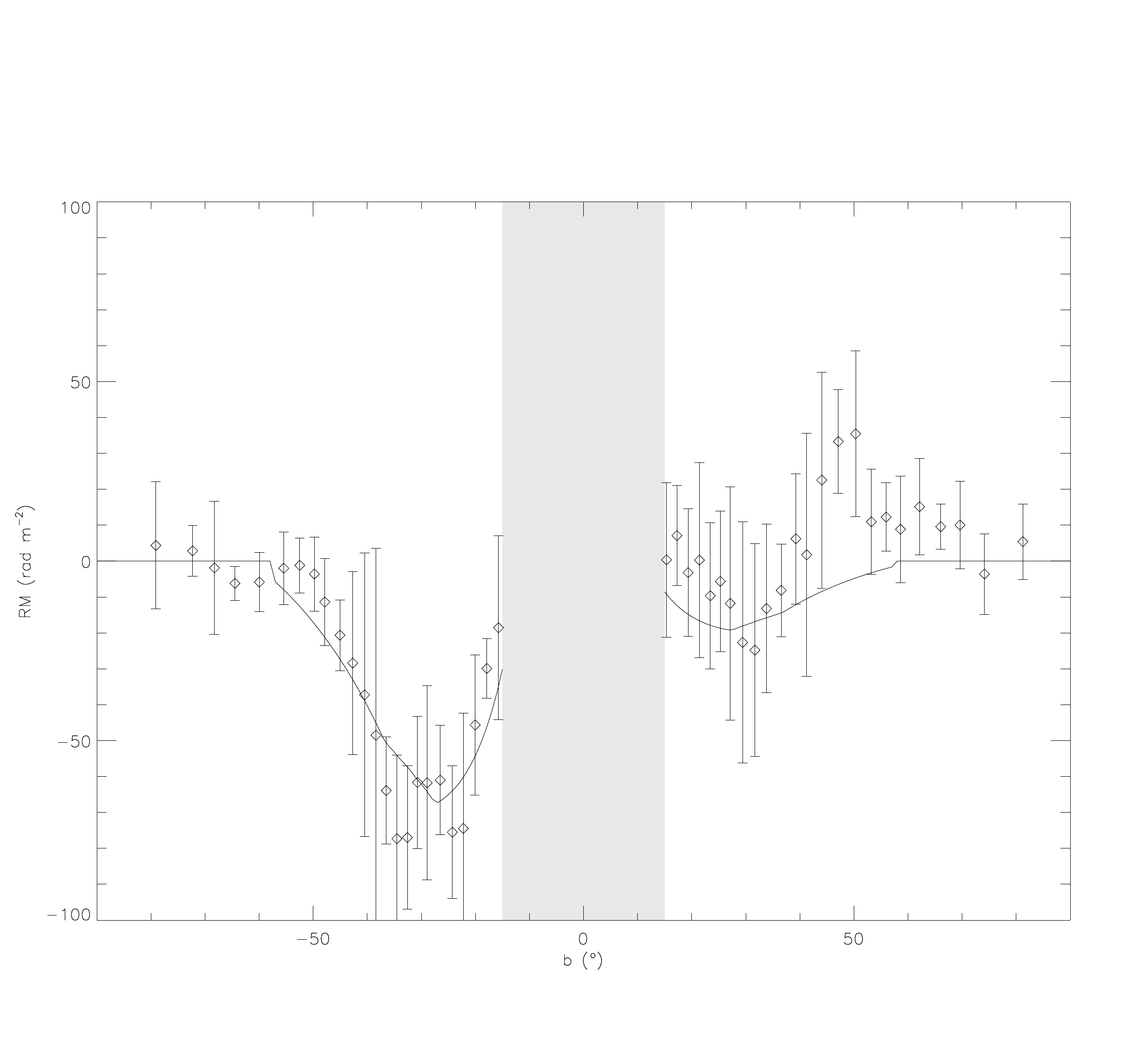}
\caption{The trend of RM as a function of Galactic latitude towards $l$=108.5$^\circ$. The predicted RM against $b$ trend from our halo field model in Section~\ref{vlasubsection:B_toy_model} (solid line) is over-plotted on the binned NVSS RMs \citep{taylor2009} denoted by diamonds. Error bars represent the weighted standard deviation of RMs within each bin. Since the Milky Way disk field dominates within 15$^\circ$ of the Galactic plane, RMs in this region are not plotted for clarity.}
\label{vlafig:B_toy_model_rm_fit_ptp}
\end{figure}

The first step to test if this simple model is realistic is to check its predictions for RM outside the VLA/CGPS coverage. We do so first by comparing the RM prediction in a strip from the south to the north Galactic pole in the same longitude range. Beyond $b$= $\pm$30$^\circ$, the \cite{taylor2009} RMs can be used to test the prediction of our model. In Figure~\ref{vlafig:B_toy_model_rm_fit_ptp}, we plot  the predicted RM as a function of latitude towards  $l$=108.5$^\circ$ from $b$=$-$90$^\circ$ to +90$^\circ$. The \cite{taylor2009} RMs are binned every 2$^\circ$ and the error bars represent the standard deviation of RM within each bin. Our model can reproduce the RM-vs-$b$ trend in the observed longitude range from the south to the north Galactic pole. We note that the disagreement between NVSS RMs and the RM prediction at  $b$=+50$^\circ$ is due to the edge of radio continuum Loop III, which could contribute at least +18 rad m$^{-2}$ to the RM of EGSs \citep{spoelstra1972}. We did not include a $\sim$ 0.3 $\mu$G vertical magnetic field component below the Galactic plane \citep{mao2010}. We note, however, that a constant and small vertical field would only shift the entire curve up or down slightly and therefore would not affect the overall shape of RM against the $b$ trend.

Next, we compare the predicted RM from our model with the observed RMs in the entire second quadrant from the south to the north Galactic pole. This is illustrated in Figure~\ref{vlafig:rm_prediction_q2}: our model can reproduce large and negative RMs below the Galactic plane (Region A). Above the Galactic plane, besides discrete regions where RMs are affected by foreground HII regions and the boundary of the radio continuum Loop III (at $b$ $\sim$ +50$^\circ$), the predicted RM distribution matches the observed EGS RMs well. The match between the NVSS RMs and our model prediction can be further improved by varying the Galacto-centric radius at which this non-zero azimuthal field exists, specifically by moving the region of non-zero magnetic field to larger Galacto-centric radii at larger Galactic longitudes. This suggests that our halo field model that can reproduce the RM-vs-$b$ trend towards 100$^\circ$$<$$l$$<$117$^\circ$ might belong to a logarithmic magnetic spiral arm in the Galactic halo. In the longitude range 100$^\circ$$<$$l$$<$117$^\circ$, the best fit parameters place a strong coherent magnetic field between Galacto-centric radii 8.8 and 10.3 kpc, roughly coinciding with the region between the local and Perseus arm (at a Galacto-centric radius of $\sim$ 9.95 kpc at  Galactic longitude of 134$^\circ$). This could be similar to the magnetic arm phenomenon seen in external galaxies such as M51 and NGC 6946, in which coherent magnetic fields are observed to be stronger in inter-arm regions than within the arms \citep{beck2007}.

\subsection{Summary and Discussion}

As shown in Sections~\ref{vlasubsection:halo_models} and \ref{vlasubsection:B_toy_model}, existing exponential and double-toroid Galactic halo field models cannot reproduce the RMs in our surveyed longitude range 100$^\circ$$<$$l$$<$117$^\circ$ at $|$$b$$|$$>$15$^\circ$. A simple double-toroid halo field also does not conform to theory and does not provide a good match to observations of the Galactic halo magnetic field. In Section~\ref{vlasubsection:B_toy_model}, we show that the presence of a coherent magnetic field within radial range 8.8-10.3 kpc at a distance 0.8-2 kpc from the Galactic plane can reproduce the RM-vs-$b$ trend within the CGPS/VLA survey region. Our model places a strong coherent field within interarm region towards the outer Galaxy. Moreover, we find that shifting the radial range of non-zero magnetic field strength to larger radii at larger longitude can provide a good match to the RM distribution in the entire second Galactic quadrant. This is suggestive of a spiral-like Galactic halo magnetic field.

\cite{beuermann1985} used the 408 MHz \cite{haslam1982} map and decomposed the Galactic synchrotron emission into a thin disk and a thick disk. The authors estimated that the Milky Way synchrotron thick disk has a scale height of 1.5 kpc at the location of the Sun and increases to 2.7 kpc beyond a Galacto-centric radius of 12 kpc. The synchrotron thick disk has comparable height to our halo magnetic field, supporting the presence of magnetic fields at distances 1-2 kpc from the Galactic mid-plane. Moreover, this thick disk exhibits spiral structure, and the magnetic field in the thick disk is modeled to align moderately with the spiral arms. We suggest that RMs in our VLA/CGPS latitude extension region could be probing the spiral magnetic field in the thick synchrotron disk.

Further insights into this proposed spiral-like thick disk field can be obtained by studying magnetic fields in Milky Way-type face-on galaxies such as M51. \cite{berkhuijsen1997} and \cite{fletcher2011} modeled M 51's  disk and halo field simultaneously, and found spiral magnetic fields (of different axial symmetry) in both its disk and its halo. The halo field has similar radial extent to the disk field with a total Faraday depth of $\sim$ 100 rad m$^{-2}$. This is in agreement with our fitted Milky Way halo field: the halo field remains substantial beyond the Solar radius, with $|$RM$|$ up to 100 rad m$^{-2}$. In M51, the ratio of the total RM through the disk to that through the halo is $\sim$ 2, which is similar to the observed value for the Milky Way. Moreover, similar to what we found in the Milky Way towards the outer Galaxy, strong coherent halo magnetic field has been found in galactocentric radii that correspond to interarm regions  in M51\citep{fletcher2011}. Other external galaxies are likely to possess spiral-like halo fields: \cite{braun2010} modeled the GHz polarized emission from a sample of SINGS galaxy with axisymmetric thick disk spiral fields with quadrupole topology from the near-side out to $\sim$ 30\% of the disk radius of these galaxies. This is further evidence that the Milky Way could possess magnetic fields in the halo at $\sim$ few kpc from the mid-plane that trace out spiral arms.

We propose that modeling of the Milky Way's halo magnetic field might require similar approaches to those used to model the magnetic field of the Galactic disk \citep[e.g.,][]{brown2007}. Sight lines towards mid- to high- Galactic latitudes could be divided into spiral arm / inter-arm regions as well as plane parallel slabs (in the $z$ direction). Each region would have a thermal electron density described by the WIM thick disk model of \cite{gaensler2008}. The strength and direction of magnetic fields within these regions can then be derived by performing a fit to the observed EGS RMs empirically. Carrying out such a fitting procedure to the entire Galactic halo is outside the scope of this paper, but we suggest that this will yield a complete picture of the Galactic halo magnetic field. Locations of magnetic field reversals as well as regions with enhanced magnetic field strength in the halo can be identified, as well as the overall axial and vertical field symmetry. A better characterization of the Milky Way halo magnetic field would be extremely useful to constrain its origin. The existing EGS RMs from \cite{taylor2009} combined with those from future surveys, such as the Polarization Sky Survey of the Universe's Magnetism (POSSUM) to be conducted with the Australian Square Kilometre Array Pathfinder (ASKAP), will provide an evenly sampled all sky RM grid for such studies.

\section{Conclusion and Future Work}
\label{vlasection:summary}
In this paper, we set out to characterize the symmetry of the disk and halo magnetic field of the Milky Way. Building on the CGPS latitude extension region, we have conducted a VLA survey of EGS RMs in the longitude range 100$^\circ$$<$$l$$<$117$^\circ$, -30$^\circ$$<$$b$$<$-3$^\circ$ and +18$^\circ$$<$$b$$<+$30$^\circ$, from which we extracted 341 reliable RMs. We have shown that the disk field in this region is symmetric about the Galactic mid-plane, consistent with the lack of warp of the ionized disk towards the outer Galaxy.  

 We have shown that towards mid Galactic latitudes in our surveyed area, RMs of EGSs are likely produced by a global halo magnetic field combined with thermal electrons in the thick WIM disk. We demonstrate the unsatisfactory predictions for the Galactic RM of existing Milky Way halo magnetic field models in the literature. Finally, we have constructed a simple halo field model with reasonable parameters that can reproduce the RM distribution in the longitude range 100$^\circ$$<$$l$$<$117$^\circ$ from $b$=$-$90$^\circ$ to $+$90$^\circ$ reasonably well. This model can be extended to explain the EGS RM distribution in the entire second Galactic quadrant. We propose future modeling of the Galactic halo magnetic field might require techniques similar to those used to model the disk magnetic field. In future work, we will make predictions of the synchrotron emission using the simple halo field model proposed in Section~\ref{vlasubsection:B_toy_model} and compare it to existing Galactic synchrotron emission survey data. We also plane to extend our simple large-scale spiral halo field model to fit to EGS/pulsar RMs at mid- to high- Galactic latitudes.

\textbf{Acknowledgements}
We thank Bob Sault for useful discussions on conducting polarization calibration of VLA data in MIRIAD. The National Radio Astronomy Observatory is a facility of the National Science Foundation operated under cooperative agreement by Associated Universities, Inc. The Canadian Galactic Plane Survey is a Canadian project with international partners and is supported by the Natural Sciences and Engineering Research Council (NSERC). The Wisconsin H-Alpha Mapper is funded by the National Science Foundation. The Southern H-Alpha Sky Survey Atlas (SHASSA) is supported by the National Science Foundation. 

B. M. G. acknowledges the support in part by an Australian Research Council Federation Fellowship (FF0561298).  M. H. acknowledges funding from the European Union's Seventh Framework Programme (FP7/2007-2013) under grant agreement number 239490 and research programme 639.042.915, which is (partly) financed by the Netherlands Organisation for Scientific Research (NWO). P. P. K acknowledges support from NSERC of Canada.

 {\it Facilities:} The Very Large Array


\clearpage

\clearpage
\clearpage

\clearpage

\clearpage

\clearpage

\clearpage
\clearpage


\clearpage

\clearpage
\LongTables
\begin{deluxetable}{llllrr}
\tablecolumns{6} 
\tablewidth{0pc} 
\tablecaption{Rotation Measures of Extragalactic Sources from VLA observations in increasing order of Galactic longitude.} \
\tablehead{   
\colhead{RA(J2000)(hms)} &     \colhead{DEC(J2000)(dms)}   &    \colhead{{\it l} ($^\circ$)} &   \colhead{{\it b} ($^\circ$)} &  \colhead{RM (rad m$^{-2}$)}   
}
\startdata 

22:49:09.60 & $+$ 42:25:12.00 & 100.06 & $-$14.96 & $-$140$\pm$ 7\\
22:23:21.60 & $+$ 50:15:36.00 & 100.16 & $-$ 5.96 & $-$ 60$\pm$21\\
22:26:38.40 & $+$ 49:28:12.00 & 100.19 & $-$ 6.90 & $-$ 59$\pm$ 6\\
22:34:38.40 & $+$ 47:22:24.00 & 100.24 & $-$ 9.37 & $-$ 54$\pm$ 5\\
22:52:12.00 & $+$ 41:43:36.00 & 100.24 & $-$15.84 & $-$100$\pm$ 5\\
22:32:21.60 & $+$ 48:19:48.00 & 100.39 & $-$ 8.37 & $-$ 82$\pm$10\\
23:11:04.80 & $+$ 34:25:48.00 & 100.42 & $-$24.03 & $-$ 36$\pm$ 4\\
23:08:26.40 & $+$ 35:37:36.00 & 100.43 & $-$22.70 & $-$ 51$\pm$ 6\\
22:44:57.60 & $+$ 44:44:36.00 & 100.50 & $-$12.56 & $-$ 47$\pm$ 8\\
22:18:28.80 & $+$ 52:04:36.00 & 100.51 & $-$ 4.02 & $-$131$\pm$ 4\\
23:12:52.80 & $+$ 33:52:12.00 & 100.54 & $-$24.68 & $-$ 71$\pm$ 1\\
22:19:19.20 & $+$ 51:58:00.00 & 100.57 & $-$ 4.16 & $-$123$\pm$11\\
22:51:36.00 & $+$ 42:42:48.00 & 100.61 & $-$14.91 & $-$119$\pm$ 6\\
23:21:43.20 & $+$ 29:55:24.00 & 100.70 & $-$29.05 & $-$100$\pm$ 7\\
22:31:24.00 & $+$ 49:18:48.00 & 100.77 & $-$ 7.44 & $-$ 60$\pm$ 8\\
22:33:09.60 & $+$ 48:52:12.00 & 100.79 & $-$ 7.97 & $-$ 74$\pm$ 3\\
22:55:57.60 & $+$ 41:30:48.00 & 100.80 & $-$16.34 & $-$101$\pm$ 5\\
22:20:50.40 & $+$ 52:08:24.00 & 100.86 & $-$ 4.15 & $-$121$\pm$ 6\\
22:38:00.00 & $+$ 47:37:24.00 & 100.87 & $-$ 9.44 & $-$ 37$\pm$12\\
23:22:24.00 & $+$ 29:57:24.00 & 100.88 & $-$29.07 & $-$105$\pm$14\\
22:47:57.60 & $+$ 44:53:60.00 & 101.06 & $-$12.68 & $-$ 51$\pm$ 7\\
23:16:43.20 & $+$ 33:14:48.00 & 101.07 & $-$25.58 & $-$100$\pm$ 4\\
22:46:40.80 & $+$ 45:23:60.00 & 101.10 & $-$12.12 & $-$ 40$\pm$ 5\\
23:12:48.00 & $+$ 35:17:24.00 & 101.16 & $-$23.37 & $-$ 59$\pm$ 4\\
22:53:55.20 & $+$ 43:04:48.00 & 101.18 & $-$14.78 & $-$ 23$\pm$12\\
23:11:40.80 & $+$ 35:55:36.00 & 101.20 & $-$22.71 & $-$ 57$\pm$ 6\\
22:48:24.00 & $+$ 45:02:36.00 & 101.20 & $-$12.58 & $-$ 35$\pm$ 1\\
23:15:16.80 & $+$ 34:23:12.00 & 101.27 & $-$24.39 & $-$ 74$\pm$14\\
22:44:57.60 & $+$ 46:27:48.00 & 101.34 & $-$11.04 & $-$ 41$\pm$ 2\\
23:19:07.20 & $+$ 32:42:24.00 & 101.35 & $-$26.26 & $-$122$\pm$ 4\\
23:03:48.00 & $+$ 39:41:36.00 & 101.37 & $-$18.63 & $-$ 37$\pm$ 9\\
22:41:55.20 & $+$ 47:37:24.00 & 101.44 & $-$ 9.78 & $-$ 73$\pm$ 6\\
22:25:26.40 & $+$ 52:15:12.00 & 101.51 & $-$ 4.44 & $-$ 90$\pm$ 7\\
22:52:07.20 & $+$ 44:29:24.00 & 101.54 & $-$13.37 & $+$ 19$\pm$26\\
22:44:04.80 & $+$ 47:16:24.00 & 101.61 & $-$10.25 & $-$ 59$\pm$ 9\\
22:34:16.80 & $+$ 50:09:36.00 & 101.61 & $-$ 6.95 & $-$101$\pm$13\\
23:15:28.80 & $+$ 35:12:60.00 & 101.66 & $-$23.68 & $-$ 64$\pm$11\\
23:17:48.00 & $+$ 34:07:12.00 & 101.68 & $-$24.84 & $-$ 94$\pm$ 5\\
23:25:02.40 & $+$ 30:25:36.00 & 101.68 & $-$28.86 & $-$125$\pm$ 9\\
23:25:02.40 & $+$ 30:25:24.00 & 101.68 & $-$28.86 & $-$134$\pm$ 4\\
22:55:31.20 & $+$ 43:44:00.00 & 101.75 & $-$14.32 & $+$ 23$\pm$16\\
23:10:16.80 & $+$ 37:52:36.00 & 101.78 & $-$20.81 & $-$ 31$\pm$13\\
23:14:12.00 & $+$ 36:16:12.00 & 101.86 & $-$22.59 & $-$ 69$\pm$19\\
23:20:16.80 & $+$ 33:25:00.00 & 101.91 & $-$25.70 & $-$ 98$\pm$11\\
22:41:50.40 & $+$ 48:43:12.00 & 101.98 & $-$ 8.80 & $-$ 87$\pm$ 4\\
22:36:55.20 & $+$ 50:10:12.00 & 101.99 & $-$ 7.14 & $-$ 73$\pm$12\\
23:28:12.00 & $+$ 29:37:48.00 & 102.07 & $-$29.86 & $-$ 77$\pm$ 1\\
22:56:02.40 & $+$ 44:17:24.00 & 102.09 & $-$13.86 & $-$ 14$\pm$ 4\\
22:50:55.20 & $+$ 46:06:12.00 & 102.11 & $-$11.83 & $-$ 14$\pm$ 1\\
23:20:38.40 & $+$ 34:10:24.00 & 102.30 & $-$25.04 & $-$ 85$\pm$16\\
23:14:43.20 & $+$ 37:07:12.00 & 102.33 & $-$21.84 & $-$ 37$\pm$14\\
22:55:31.20 & $+$ 45:01:48.00 & 102.34 & $-$13.16 & $-$  3$\pm$ 3\\
22:52:28.80 & $+$ 46:18:60.00 & 102.44 & $-$11.77 & $-$ 21$\pm$12\\
23:17:33.60 & $+$ 36:15:00.00 & 102.54 & $-$22.87 & $-$ 78$\pm$ 9\\
23:01:04.80 & $+$ 43:23:12.00 & 102.54 & $-$15.07 & $-$ 28$\pm$ 4\\
23:07:52.80 & $+$ 40:42:36.00 & 102.56 & $-$18.04 & $-$ 35$\pm$ 4\\
23:25:45.60 & $+$ 32:12:12.00 & 102.59 & $-$27.25 & $-$141$\pm$ 2\\
23:19:00.00 & $+$ 35:56:12.00 & 102.70 & $-$23.27 & $-$ 57$\pm$ 8\\
23:18:52.80 & $+$ 36:05:48.00 & 102.74 & $-$23.13 & $-$ 74$\pm$ 8\\
23:07:24.00 & $+$ 41:32:36.00 & 102.83 & $-$17.24 & $-$ 29$\pm$ 8\\
23:14:00.00 & $+$ 39:01:48.00 & 102.99 & $-$20.04 & $-$ 70$\pm$11\\
23:17:07.20 & $+$ 37:43:48.00 & 103.05 & $-$21.48 & $-$ 65$\pm$12\\
23:01:19.20 & $+$ 44:38:12.00 & 103.13 & $-$13.95 & $+$ 29$\pm$19\\
22:57:31.20 & $+$ 46:04:48.00 & 103.13 & $-$12.38 & $-$  2$\pm$23\\
22:35:36.00 & $+$ 52:50:36.00 & 103.14 & $-$ 4.73 & $-$ 80$\pm$ 3\\
22:40:19.20 & $+$ 51:32:48.00 & 103.14 & $-$ 6.21 & $-$ 74$\pm$ 2\\
22:55:12.00 & $+$ 46:59:60.00 & 103.18 & $-$11.38 & $-$ 17$\pm$ 6\\
23:02:12.00 & $+$ 44:48:24.00 & 103.35 & $-$13.86 & $-$ 14$\pm$18\\
23:06:31.20 & $+$ 43:10:24.00 & 103.38 & $-$15.67 & $+$  2$\pm$ 6\\
23:04:40.80 & $+$ 44:01:12.00 & 103.42 & $-$14.77 & $-$ 13$\pm$ 4\\
23:27:31.20 & $+$ 33:17:36.00 & 103.42 & $-$26.37 & $-$ 67$\pm$ 9\\
22:49:28.80 & $+$ 49:24:36.00 & 103.42 & $-$ 8.79 & $-$ 45$\pm$28\\
22:45:26.40 & $+$ 50:41:24.00 & 103.43 & $-$ 7.35 & $-$ 61$\pm$17\\
23:22:31.20 & $+$ 36:00:36.00 & 103.44 & $-$23.48 & $-$ 69$\pm$ 8\\
22:36:55.20 & $+$ 53:07:12.00 & 103.45 & $-$ 4.59 & $-$ 26$\pm$ 8\\
22:51:07.20 & $+$ 49:00:36.00 & 103.48 & $-$ 9.27 & $-$ 50$\pm$16\\
22:46:57.60 & $+$ 50:21:24.00 & 103.49 & $-$ 7.76 & $-$ 78$\pm$14\\
23:28:14.40 & $+$ 33:08:48.00 & 103.51 & $-$26.58 & $-$ 86$\pm$23\\
22:57:04.80 & $+$ 47:05:24.00 & 103.52 & $-$11.43 & $-$ 15$\pm$10\\
23:07:12.00 & $+$ 43:23:24.00 & 103.58 & $-$15.53 & $-$ 19$\pm$10\\
23:21:04.80 & $+$ 37:18:24.00 & 103.67 & $-$22.16 & $-$ 46$\pm$ 8\\
22:50:26.40 & $+$ 49:48:00.00 & 103.74 & $-$ 8.50 & $-$ 42$\pm$ 3\\
22:52:04.80 & $+$ 49:16:24.00 & 103.74 & $-$ 9.09 & $-$ 47$\pm$ 7\\
22:53:52.80 & $+$ 49:17:48.00 & 104.01 & $-$ 9.21 & $-$ 40$\pm$14\\
23:01:31.20 & $+$ 46:36:12.00 & 104.01 & $-$12.19 & $-$  9$\pm$13\\
23:27:38.40 & $+$ 34:50:12.00 & 104.05 & $-$24.94 & $-$ 80$\pm$10\\
23:12:09.60 & $+$ 42:49:12.00 & 104.21 & $-$16.42 & $-$  9$\pm$ 3\\
23:24:00.00 & $+$ 37:19:12.00 & 104.26 & $-$22.35 & $-$ 87$\pm$14\\
23:02:26.40 & $+$ 46:53:24.00 & 104.28 & $-$11.99 & $-$ 29$\pm$ 6\\
23:19:21.60 & $+$ 40:10:12.00 & 104.47 & $-$19.38 & $-$ 67$\pm$19\\
23:24:43.20 & $+$ 37:38:48.00 & 104.53 & $-$22.10 & $-$ 61$\pm$ 8\\
23:07:48.00 & $+$ 45:37:48.00 & 104.61 & $-$13.52 & $+$ 12$\pm$ 7\\
22:53:31.20 & $+$ 50:57:24.00 & 104.70 & $-$ 7.71 & $-$ 38$\pm$ 9\\
22:42:16.80 & $+$ 54:16:48.00 & 104.71 & $-$ 3.96 & $-$112$\pm$24\\
23:20:36.00 & $+$ 40:22:36.00 & 104.78 & $-$19.28 & $-$ 54$\pm$ 8\\
23:16:24.00 & $+$ 42:26:24.00 & 104.81 & $-$17.06 & $-$ 22$\pm$11\\
23:27:36.00 & $+$ 37:01:24.00 & 104.87 & $-$22.88 & $-$ 93$\pm$16\\
23:16:02.40 & $+$ 43:15:36.00 & 105.06 & $-$16.29 & $-$ 14$\pm$12\\
22:48:28.80 & $+$ 53:25:12.00 & 105.12 & $-$ 5.14 & $-$ 59$\pm$ 3\\
22:59:45.60 & $+$ 49:53:12.00 & 105.15 & $-$ 9.08 & $-$ 49$\pm$ 9\\
23:25:40.80 & $+$ 39:06:36.00 & 105.26 & $-$20.81 & $-$ 47$\pm$11\\
23:15:07.20 & $+$ 44:13:12.00 & 105.29 & $-$15.31 & $+$  8$\pm$23\\
23:11:45.60 & $+$ 45:43:12.00 & 105.31 & $-$13.70 & $-$394$\pm$37\\
22:46:12.00 & $+$ 54:51:36.00 & 105.49 & $-$ 3.71 & $-$103$\pm$ 5\\
23:13:02.40 & $+$ 45:46:12.00 & 105.54 & $-$13.75 & $-$  3$\pm$ 7\\
23:01:14.40 & $+$ 50:34:24.00 & 105.65 & $-$ 8.55 & $-$ 52$\pm$12\\
23:20:00.00 & $+$ 43:17:12.00 & 105.78 & $-$16.50 & $-$ 33$\pm$15\\
18:50:16.80 & $+$ 74:40:60.00 & 105.80 & $+$26.17 & $-$ 37$\pm$ 6\\
23:21:14.40 & $+$ 42:49:48.00 & 105.82 & $-$17.04 & $-$ 42$\pm$ 5\\
23:26:14.40 & $+$ 40:33:24.00 & 105.91 & $-$19.48 & $-$ 69$\pm$ 4\\
23:28:00.00 & $+$ 39:55:00.00 & 106.02 & $-$20.19 & $-$ 48$\pm$ 6\\
18:09:21.60 & $+$ 75:03:36.00 & 106.11 & $+$28.85 & $-$ 61$\pm$ 4\\
23:21:57.60 & $+$ 43:13:12.00 & 106.11 & $-$16.70 & $-$ 56$\pm$ 6\\
23:31:50.40 & $+$ 38:06:36.00 & 106.12 & $-$22.16 & $-$ 70$\pm$ 2\\
23:12:36.00 & $+$ 47:29:36.00 & 106.14 & $-$12.13 & $-$ 38$\pm$ 5\\
23:11:14.40 & $+$ 48:07:36.00 & 106.18 & $-$11.45 & $-$ 16$\pm$ 2\\
23:13:33.60 & $+$ 47:28:48.00 & 106.28 & $-$12.21 & $-$  9$\pm$ 8\\
23:07:07.20 & $+$ 50:07:24.00 & 106.33 & $-$ 9.36 & $-$ 50$\pm$ 6\\
22:50:43.20 & $+$ 55:50:00.00 & 106.51 & $-$ 3.14 & $-$ 89$\pm$ 9\\
18:49:52.80 & $+$ 75:19:24.00 & 106.52 & $+$26.27 & $-$ 61$\pm$ 3\\
23:16:38.40 & $+$ 46:51:12.00 & 106.54 & $-$12.98 & $+$  1$\pm$ 5\\
19:37:04.80 & $+$ 74:40:24.00 & 106.57 & $+$23.16 & $-$ 27$\pm$ 7\\
19:13:55.20 & $+$ 75:08:24.00 & 106.61 & $+$24.73 & $-$ 50$\pm$19\\
23:14:21.60 & $+$ 48:00:36.00 & 106.62 & $-$11.77 & $+$  5$\pm$ 3\\
23:13:00.00 & $+$ 48:40:48.00 & 106.66 & $-$11.07 & $-$ 17$\pm$ 4\\
19:20:24.00 & $+$ 75:05:36.00 & 106.67 & $+$24.32 & $-$113$\pm$13\\
22:55:40.80 & $+$ 54:45:60.00 & 106.67 & $-$ 4.42 & $-$129$\pm$14\\
23:22:19.20 & $+$ 44:45:12.00 & 106.73 & $-$15.29 & $-$ 38$\pm$ 8\\
23:38:21.60 & $+$ 36:06:48.00 & 106.78 & $-$24.46 & $-$ 65$\pm$11\\
23:17:04.80 & $+$ 47:22:24.00 & 106.81 & $-$12.52 & $-$ 11$\pm$ 8\\
23:45:07.20 & $+$ 31:45:48.00 & 106.82 & $-$29.04 & $-$ 64$\pm$ 8\\
23:35:28.80 & $+$ 38:13:48.00 & 106.89 & $-$22.27 & $-$ 55$\pm$ 6\\
18:08:14.40 & $+$ 75:49:12.00 & 106.99 & $+$28.88 & $-$ 66$\pm$ 3\\
23:24:02.40 & $+$ 44:55:12.00 & 107.08 & $-$15.24 & $-$ 23$\pm$ 9\\
23:31:07.20 & $+$ 41:16:48.00 & 107.08 & $-$19.12 & $-$ 36$\pm$ 4\\
23:01:55.20 & $+$ 53:47:00.00 & 107.09 & $-$ 5.67 & $-$ 49$\pm$12\\
23:39:28.80 & $+$ 36:20:12.00 & 107.09 & $-$24.31 & $-$ 86$\pm$10\\
23:14:31.20 & $+$ 49:17:12.00 & 107.13 & $-$10.59 & $-$ 27$\pm$ 7\\
23:40:45.60 & $+$ 35:53:60.00 & 107.21 & $-$24.82 & $-$128$\pm$ 9\\
23:34:28.80 & $+$ 39:46:36.00 & 107.22 & $-$20.73 & $-$ 51$\pm$10\\
23:21:24.00 & $+$ 46:34:12.00 & 107.23 & $-$13.53 & $-$ 13$\pm$11\\
18:30:09.60 & $+$ 76:06:36.00 & 107.30 & $+$27.55 & $-$ 66$\pm$ 8\\
23:20:57.60 & $+$ 47:05:48.00 & 107.34 & $-$13.02 & $-$ 11$\pm$ 6\\
23:06:57.60 & $+$ 52:45:60.00 & 107.36 & $-$ 6.93 & $-$ 37$\pm$20\\
23:36:52.80 & $+$ 38:50:00.00 & 107.38 & $-$21.77 & $-$ 60$\pm$17\\
23:29:43.20 & $+$ 43:04:36.00 & 107.44 & $-$17.33 & $-$ 34$\pm$ 9\\
23:05:28.80 & $+$ 53:30:24.00 & 107.46 & $-$ 6.15 & $-$113$\pm$13\\
23:18:26.40 & $+$ 48:37:48.00 & 107.49 & $-$11.44 & $-$  7$\pm$ 6\\
23:43:50.40 & $+$ 34:49:24.00 & 107.54 & $-$26.01 & $-$ 63$\pm$ 8\\
23:06:16.80 & $+$ 53:31:24.00 & 107.57 & $-$ 6.19 & $-$ 75$\pm$11\\
23:41:12.00 & $+$ 36:43:36.00 & 107.58 & $-$24.04 & $-$ 43$\pm$ 9\\
23:38:52.80 & $+$ 38:23:60.00 & 107.64 & $-$22.32 & $-$ 58$\pm$ 4\\
20:17:14.40 & $+$ 74:40:00.00 & 107.69 & $+$20.72 & $-$ 27$\pm$ 5\\
17:55:55.20 & $+$ 76:27:00.00 & 107.79 & $+$29.56 & $-$ 25$\pm$ 4\\
23:14:19.20 & $+$ 51:10:12.00 & 107.81 & $-$ 8.83 & $-$101$\pm$13\\
23:48:50.40 & $+$ 32:08:12.00 & 107.81 & $-$28.88 & $-$ 70$\pm$ 3\\
23:08:45.60 & $+$ 53:21:00.00 & 107.85 & $-$ 6.49 & $-$ 80$\pm$14\\
23:38:09.60 & $+$ 39:33:48.00 & 107.87 & $-$21.16 & $-$ 54$\pm$ 7\\
23:17:12.00 & $+$ 50:23:12.00 & 107.95 & $-$ 9.72 & $-$ 10$\pm$11\\
23:26:57.60 & $+$ 46:19:24.00 & 108.07 & $-$14.09 & $+$ 12$\pm$10\\
23:11:14.40 & $+$ 53:10:12.00 & 108.12 & $-$ 6.79 & $-$ 78$\pm$10\\
19:34:36.00 & $+$ 76:17:12.00 & 108.20 & $+$23.80 & $-$ 32$\pm$14\\
18:45:02.40 & $+$ 76:52:36.00 & 108.21 & $+$26.73 & $-$ 84$\pm$ 2\\
23:15:07.20 & $+$ 52:06:24.00 & 108.27 & $-$ 8.01 & $-$ 65$\pm$ 7\\
19:04:26.40 & $+$ 76:47:12.00 & 108.27 & $+$25.62 & $-$ 58$\pm$11\\
23:33:21.60 & $+$ 43:46:00.00 & 108.33 & $-$16.87 & $-$ 35$\pm$ 6\\
23:46:33.60 & $+$ 35:28:00.00 & 108.33 & $-$25.56 & $-$ 65$\pm$ 4\\
23:32:04.80 & $+$ 44:30:12.00 & 108.34 & $-$16.11 & $-$ 14$\pm$ 4\\
23:23:07.20 & $+$ 49:05:48.00 & 108.40 & $-$11.27 & $-$ 61$\pm$30\\
23:41:14.40 & $+$ 39:18:24.00 & 108.40 & $-$21.57 & $-$ 62$\pm$12\\
23:25:45.60 & $+$ 48:06:12.00 & 108.47 & $-$12.34 & $-$ 44$\pm$11\\
23:16:21.60 & $+$ 52:19:24.00 & 108.53 & $-$ 7.88 & $-$ 81$\pm$ 9\\
23:35:48.00 & $+$ 43:06:48.00 & 108.55 & $-$17.65 & $-$ 33$\pm$11\\
23:51:43.20 & $+$ 32:21:48.00 & 108.55 & $-$28.83 & $-$ 68$\pm$10\\
23:31:48.00 & $+$ 45:22:36.00 & 108.58 & $-$15.26 & $-$ 12$\pm$ 9\\
18:43:14.40 & $+$ 77:15:24.00 & 108.62 & $+$26.86 & $-$ 75$\pm$ 9\\
23:37:24.00 & $+$ 42:28:12.00 & 108.65 & $-$18.34 & $-$ 43$\pm$ 8\\
19:32:04.80 & $+$ 76:46:36.00 & 108.66 & $+$24.08 & $-$ 41$\pm$10\\
23:27:26.40 & $+$ 48:02:12.00 & 108.73 & $-$12.49 & $-$ 22$\pm$ 4\\
23:25:31.20 & $+$ 49:01:48.00 & 108.74 & $-$11.47 & $-$ 47$\pm$19\\
20:38:07.20 & $+$ 75:05:48.00 & 108.80 & $+$19.75 & $-$ 18$\pm$ 7\\
23:48:21.60 & $+$ 35:54:12.00 & 108.85 & $-$25.23 & $-$ 82$\pm$ 6\\
23:53:52.80 & $+$ 32:09:00.00 & 108.99 & $-$29.15 & $-$ 54$\pm$ 9\\
23:25:38.40 & $+$ 50:03:12.00 & 109.12 & $-$10.50 & $-$ 73$\pm$11\\
23:39:19.20 & $+$ 42:57:00.00 & 109.15 & $-$17.98 & $-$ 65$\pm$20\\
23:55:43.20 & $+$ 31:25:48.00 & 109.22 & $-$29.95 & $-$ 98$\pm$19\\
23:52:38.40 & $+$ 34:10:12.00 & 109.28 & $-$27.14 & $-$ 48$\pm$23\\
20:22:36.00 & $+$ 76:11:24.00 & 109.29 & $+$21.13 & $-$ 34$\pm$ 3\\
23:42:48.00 & $+$ 41:16:12.00 & 109.31 & $-$19.77 & $-$512$\pm$32\\
23:28:50.40 & $+$ 49:19:36.00 & 109.37 & $-$11.36 & $-$ 76$\pm$ 8\\
23:49:43.20 & $+$ 36:59:12.00 & 109.45 & $-$24.26 & $-$112$\pm$14\\
23:53:19.20 & $+$ 34:17:36.00 & 109.48 & $-$27.06 & $-$ 90$\pm$15\\
18:07:40.80 & $+$ 78:04:48.00 & 109.56 & $+$28.75 & $-$ 64$\pm$12\\
20:13:57.60 & $+$ 76:46:36.00 & 109.60 & $+$21.85 & $-$ 35$\pm$10\\
20:56:43.20 & $+$ 75:14:12.00 & 109.64 & $+$18.85 & $-$ 35$\pm$ 2\\
23:26:04.80 & $+$ 51:40:48.00 & 109.73 & $-$ 8.99 & $-$155$\pm$12\\
23:49:21.60 & $+$ 38:49:36.00 & 109.89 & $-$22.47 & $-$103$\pm$19\\
23:56:48.00 & $+$ 33:04:36.00 & 109.93 & $-$28.42 & $-$ 84$\pm$ 8\\
23:40:07.20 & $+$ 45:34:36.00 & 110.08 & $-$15.52 & $-$ 62$\pm$10\\
20:12:48.00 & $+$ 77:19:60.00 & 110.09 & $+$22.15 & $-$ 44$\pm$ 4\\
23:42:12.00 & $+$ 44:41:00.00 & 110.19 & $-$16.47 & $-$ 66$\pm$17\\
23:26:31.20 & $+$ 52:57:24.00 & 110.22 & $-$ 7.80 & $-$123$\pm$22\\
23:39:09.60 & $+$ 46:41:12.00 & 110.24 & $-$14.40 & $-$ 45$\pm$16\\
20:45:40.80 & $+$ 76:25:00.00 & 110.25 & $+$20.08 & $-$ 12$\pm$ 2\\
23:59:50.40 & $+$ 31:40:48.00 & 110.27 & $-$29.93 & $-$ 96$\pm$15\\
23:34:43.20 & $+$ 49:15:36.00 & 110.28 & $-$11.72 & $-$ 35$\pm$ 7\\
23:30:07.20 & $+$ 51:54:48.00 & 110.40 & $-$ 8.97 & $-$514$\pm$25\\
23:41:36.00 & $+$ 45:49:24.00 & 110.41 & $-$15.34 & $-$ 43$\pm$12\\
23:41:36.00 & $+$ 46:03:12.00 & 110.48 & $-$15.11 & $-$ 50$\pm$ 4\\
23:48:45.60 & $+$ 41:27:48.00 & 110.50 & $-$19.90 & $-$ 61$\pm$16\\
21:18:14.40 & $+$ 75:12:12.00 & 110.51 & $+$17.76 & $+$  7$\pm$ 1\\
21:27:48.00 & $+$ 74:50:48.00 & 110.67 & $+$17.06 & $-$ 12$\pm$12\\
23:26:24.00 & $+$ 54:49:36.00 & 110.80 & $-$ 6.03 & $-$ 98$\pm$12\\
23:36:26.40 & $+$ 50:08:24.00 & 110.82 & $-$10.95 & $-$ 60$\pm$20\\
23:44:36.00 & $+$ 46:04:12.00 & 111.00 & $-$15.25 & $-$ 46$\pm$30\\
21:22:57.60 & $+$ 75:37:24.00 & 111.05 & $+$17.82 & $+$ 25$\pm$ 3\\
23:46:24.00 & $+$ 45:19:12.00 & 111.11 & $-$16.05 & $-$438$\pm$16\\
19:43:38.40 & $+$ 78:58:36.00 & 111.15 & $+$24.15 & $-$ 16$\pm$ 1\\
20:12:40.80 & $+$ 78:28:12.00 & 111.22 & $+$22.65 & $+$  8$\pm$ 7\\
23:40:52.80 & $+$ 49:36:24.00 & 111.36 & $-$11.68 & $-$482$\pm$27\\
23:35:50.40 & $+$ 52:15:12.00 & 111.36 & $-$ 8.91 & $+$336$\pm$11\\
23:50:38.40 & $+$ 43:40:12.00 & 111.44 & $-$17.84 & $-$ 60$\pm$10\\
20:08:52.80 & $+$ 78:55:00.00 & 111.57 & $+$23.00 & $+$ 10$\pm$12\\
21:02:36.00 & $+$ 77:15:12.00 & 111.57 & $+$19.78 & $-$ 14$\pm$13\\
23:49:21.60 & $+$ 45:35:36.00 & 111.70 & $-$15.92 & $-$ 50$\pm$24\\
18:53:09.60 & $+$ 80:01:00.00 & 111.77 & $+$26.61 & $-$ 15$\pm$29\\
23:44:12.00 & $+$ 49:07:48.00 & 111.77 & $-$12.28 & $-$ 41$\pm$21\\
00:01:55.20 & $+$ 36:22:48.00 & 111.92 & $-$25.44 & $-$104$\pm$13\\
23:47:07.20 & $+$ 48:01:36.00 & 111.95 & $-$13.47 & $-$ 33$\pm$17\\
21:19:57.60 & $+$ 76:57:12.00 & 111.98 & $+$18.85 & $+$  2$\pm$ 9\\
23:50:16.80 & $+$ 46:06:36.00 & 112.00 & $-$15.46 & $-$ 46$\pm$18\\
17:43:40.80 & $+$ 80:04:00.00 & 112.00 & $+$29.60 & $-$ 42$\pm$ 4\\
19:34:19.20 & $+$ 79:56:00.00 & 112.03 & $+$24.84 & $+$ 14$\pm$ 3\\
23:49:04.80 & $+$ 47:21:12.00 & 112.11 & $-$14.20 & $-$ 16$\pm$15\\
23:38:52.80 & $+$ 53:37:60.00 & 112.19 & $-$ 7.73 & $-$108$\pm$20\\
19:12:24.00 & $+$ 80:26:36.00 & 112.34 & $+$25.86 & $-$ 20$\pm$ 5\\
23:37:45.60 & $+$ 55:51:12.00 & 112.67 & $-$ 5.55 & $-$109$\pm$32\\
23:53:28.80 & $+$ 46:51:24.00 & 112.73 & $-$14.87 & $-$ 52$\pm$22\\
00:01:28.80 & $+$ 41:04:24.00 & 112.89 & $-$20.82 & $-$ 74$\pm$14\\
20:58:43.20 & $+$ 79:05:00.00 & 113.01 & $+$21.03 & $+$ 14$\pm$11\\
00:00:38.40 & $+$ 43:57:48.00 & 113.36 & $-$17.97 & $-$ 70$\pm$ 8\\
23:52:12.00 & $+$ 50:26:24.00 & 113.37 & $-$11.33 & $-$  1$\pm$22\\
21:57:12.00 & $+$ 76:46:00.00 & 113.43 & $+$17.24 & $-$ 19$\pm$ 5\\
18:11:48.00 & $+$ 81:30:24.00 & 113.43 & $+$28.27 & $+$ 26$\pm$ 4\\
21:54:07.20 & $+$ 77:04:48.00 & 113.49 & $+$17.57 & $+$ 10$\pm$ 4\\
23:51:00.00 & $+$ 51:49:24.00 & 113.52 & $-$ 9.94 & $-$ 55$\pm$21\\
00:05:07.20 & $+$ 40:57:60.00 & 113.58 & $-$21.08 & $-$ 72$\pm$23\\
23:47:28.80 & $+$ 54:32:12.00 & 113.67 & $-$ 7.17 & $-$ 38$\pm$21\\
00:11:52.80 & $+$ 34:16:36.00 & 113.69 & $-$27.88 & $-$ 73$\pm$13\\
18:45:33.60 & $+$ 81:50:24.00 & 113.76 & $+$27.04 & $+$ 15$\pm$ 1\\
00:03:19.20 & $+$ 43:34:36.00 & 113.77 & $-$18.44 & $-$ 51$\pm$ 8\\
17:51:00.00 & $+$ 81:47:60.00 & 113.87 & $+$28.98 & $-$ 17$\pm$ 4\\
00:12:40.80 & $+$ 35:39:12.00 & 114.14 & $-$26.55 & $-$ 76$\pm$33\\
00:14:45.60 & $+$ 33:40:36.00 & 114.24 & $-$28.59 & $-$ 65$\pm$18\\
00:04:55.20 & $+$ 44:27:48.00 & 114.25 & $-$17.63 & $-$ 59$\pm$12\\
21:09:16.80 & $+$ 80:20:24.00 & 114.40 & $+$21.42 & $+$ 21$\pm$14\\
00:06:14.40 & $+$ 44:04:24.00 & 114.42 & $-$18.04 & $-$ 41$\pm$ 8\\
23:51:40.80 & $+$ 55:32:60.00 & 114.49 & $-$ 6.35 & $-$ 85$\pm$17\\
00:15:14.40 & $+$ 35:03:48.00 & 114.60 & $-$27.23 & $-$ 75$\pm$ 6\\
21:15:50.40 & $+$ 80:29:00.00 & 114.69 & $+$21.29 & $+$ 31$\pm$ 6\\
19:38:36.00 & $+$ 82:33:36.00 & 114.87 & $+$25.38 & $+$ 14$\pm$28\\
00:15:31.20 & $+$ 36:12:36.00 & 114.87 & $-$26.11 & $-$ 82$\pm$12\\
23:57:28.80 & $+$ 53:31:24.00 & 114.87 & $-$ 8.50 & $-$ 53$\pm$25\\
23:59:43.20 & $+$ 52:37:36.00 & 115.02 & $-$ 9.45 & $+$ 10$\pm$10\\
00:12:55.20 & $+$ 40:32:24.00 & 115.07 & $-$21.73 & $-$ 80$\pm$12\\
23:56:38.40 & $+$ 55:18:24.00 & 115.13 & $-$ 6.73 & $-$ 55$\pm$15\\
00:11:07.20 & $+$ 43:15:48.00 & 115.19 & $-$19.00 & $-$ 15$\pm$10\\
23:57:00.00 & $+$ 55:41:48.00 & 115.26 & $-$ 6.37 & $-$ 69$\pm$ 5\\
19:45:28.80 & $+$ 82:57:00.00 & 115.35 & $+$25.28 & $+$  8$\pm$ 6\\
00:18:52.80 & $+$ 34:38:12.00 & 115.37 & $-$27.75 & $-$ 51$\pm$11\\
23:55:33.60 & $+$ 57:28:60.00 & 115.45 & $-$ 4.59 & $-$ 66$\pm$11\\
21:10:07.20 & $+$ 81:36:36.00 & 115.49 & $+$22.16 & $-$  5$\pm$10\\
00:12:14.40 & $+$ 43:58:36.00 & 115.52 & $-$18.32 & $-$ 49$\pm$10\\
21:25:04.80 & $+$ 81:15:36.00 & 115.57 & $+$21.51 & $+$ 22$\pm$36\\
17:09:16.80 & $+$ 83:00:12.00 & 115.65 & $+$29.95 & $-$ 48$\pm$ 5\\
00:06:07.20 & $+$ 50:49:36.00 & 115.66 & $-$11.40 & $+$ 37$\pm$ 4\\
22:17:19.20 & $+$ 78:58:24.00 & 115.72 & $+$18.32 & $-$ 21$\pm$ 8\\
17:05:07.20 & $+$ 83:16:12.00 & 115.98 & $+$29.97 & $-$ 40$\pm$ 3\\
22:15:57.60 & $+$ 79:29:12.00 & 115.99 & $+$18.78 & $-$ 26$\pm$ 6\\
00:17:33.60 & $+$ 40:47:12.00 & 116.04 & $-$21.63 & $-$ 74$\pm$ 5\\
00:10:40.80 & $+$ 48:47:48.00 & 116.05 & $-$13.52 & $-$  9$\pm$ 9\\
00:12:40.80 & $+$ 47:04:48.00 & 116.11 & $-$15.29 & $-$ 18$\pm$ 8\\
00:11:00.00 & $+$ 49:04:24.00 & 116.15 & $-$13.26 & $-$  8$\pm$ 4\\
00:05:50.40 & $+$ 54:00:00.00 & 116.19 & $-$ 8.28 & $-$ 49$\pm$17\\
22:18:24.00 & $+$ 79:41:24.00 & 116.21 & $+$18.88 & $-$  8$\pm$ 3\\
00:05:16.80 & $+$ 55:17:12.00 & 116.34 & $-$ 6.99 & $-$ 53$\pm$14\\
00:07:24.00 & $+$ 53:32:60.00 & 116.34 & $-$ 8.77 & $-$  2$\pm$ 5\\
00:11:52.80 & $+$ 49:31:00.00 & 116.37 & $-$12.84 & $+$ 18$\pm$12\\
00:04:28.80 & $+$ 56:14:12.00 & 116.40 & $-$ 6.04 & $-$ 56$\pm$ 7\\
21:27:50.40 & $+$ 82:13:24.00 & 116.42 & $+$22.07 & $+$ 16$\pm$ 5\\
00:22:12.00 & $+$ 36:57:60.00 & 116.46 & $-$25.55 & $-$104$\pm$ 8\\
00:24:43.20 & $+$ 32:51:12.00 & 116.48 & $-$29.68 & $-$ 46$\pm$19\\
18:31:43.20 & $+$ 84:16:60.00 & 116.49 & $+$27.47 & $+$  4$\pm$ 7\\
00:11:36.00 & $+$ 50:41:60.00 & 116.51 & $-$11.68 & $+$ 18$\pm$ 6\\
21:09:38.40 & $+$ 82:55:48.00 & 116.60 & $+$22.97 & $-$  8$\pm$10\\
00:15:16.80 & $+$ 47:36:48.00 & 116.64 & $-$14.82 & $-$ 37$\pm$ 5\\
00:25:19.20 & $+$ 33:00:00.00 & 116.64 & $-$29.54 & $-$ 50$\pm$ 8\\
00:16:43.20 & $+$ 46:08:48.00 & 116.68 & $-$16.31 & $-$ 20$\pm$ 8\\
20:23:33.60 & $+$ 83:54:12.00 & 116.77 & $+$24.61 & $-$ 15$\pm$18\\
00:11:14.40 & $+$ 53:39:12.00 & 116.92 & $-$ 8.75 & $-$ 21$\pm$ 6\\
00:15:40.80 & $+$ 49:07:48.00 & 116.93 & $-$13.33 & $-$ 32$\pm$ 3\\
00:19:14.40 & $+$ 44:49:36.00 & 116.95 & $-$17.67 & $-$ 33$\pm$ 6\\

\enddata
 \label{vlatable:rm_catalogue}
\end{deluxetable}

{\singlespace
\centering
\begin{deluxetable}{lrrrrrr}
\tabletypesize{\footnotesize}
\tablecolumns{7} 
\tablewidth{0pc} 
\tablecaption{Double-Torus Halo Magnetic Field Parameters} 
\tablehead{   
\colhead {Reference} & \colhead{$B^H_0$ ($\mu$ G) } & \colhead{$r^H_0$ (kpc)}   & \colhead{$z^H_0$ (kpc)} &  \colhead{$z^H_1$ (kpc) for $|z|$$<$$z^H_0$} &  \colhead{$z^H_1$ (kpc) for $|z|$$>$$z^H_0$} & \colhead{Model type}
}
\startdata
 \cite{sun2008} & 10   & 4  & 1.5 & 0.2 & 0.4  & Qualitative \\
 \cite{jansson2009}  & 4.9  & 18 &1.4  & 0.12  &  8.5  & Fitted\tablenotemark{a}\\
 \cite{sun2010} & 2  & 4 &  1.5 & 0.2  &  4  & Qualitative \\
 \multirow{2}{*}{\cite{pshirkov2011}} & 4   for z$>$0 & \multirow{2}{*}{6 } & \multirow{2}{*}{1.3 } & \multirow{2}{*}{0.25 } &\multirow{2}{*}{0.4 } & \multirow{2}{*}{Benchmark\tablenotemark{b}} \\
 & 2  for z$<$0 & & & & \\
\enddata 
\label{vlatable:ps_field_form}
\tablenotetext{a}{These parameters were found by fitting to EGS RMs and diffuse polarized synchrotron emission.}
\tablenotetext{b}{These parameters were chosen based on model fits to EGS RMs.}
\end{deluxetable} 
}

\clearpage

\end{document}